\documentclass[%
 aip,
 jcp,%
 amsmath,amssymb,
reprint,%
longbibliography
]{revtex4-1}

\usepackage[printwatermark]{xwatermark}
\usepackage{graphicx}
\usepackage{epstopdf}
\usepackage{dcolumn}
\usepackage{bm}
\usepackage{hyperref}
\usepackage{comment}
\usepackage{color}

\newcommand{\im}{\mathrm{i}}
\newcommand{\dif}{\mathrm{d}}

\newcommand{\e}{\mathrm{e}}

\begin{document}

\title{Nonequilibrium steady-state picture of incoherent light-induced excitation harvesting}

\author{Veljko Jankovi\'{c}}
\email{veljko.jankovic@ipb.ac.rs}
\affiliation{Scientific Computing Laboratory, Center for the Study of Complex Systems, Institute of Physics Belgrade, University of Belgrade, Pregrevica 118, 11080 Belgrade, Serbia}

\author{Tom\'{a}\v{s} Man\v{c}al}%
 \email{mancal@karlov.mff.cuni.cz}
\affiliation{
 Faculty of Mathematics and Physics, Charles University, Ke Karlovu 5, 121 16 Prague 2, Czech Republic
}

\begin{abstract}
We formulate a comprehensive theoretical description of excitation harvesting in molecular aggregates photoexcited by weak incoherent radiation. An efficient numerical scheme that respects the continuity equation for excitation fluxes is developed to compute the nonequilibrium steady state (NESS) arising from the interplay between excitation generation, excitation relaxation, dephasing, trapping at the load, and recombination. The NESS is most conveniently described in the so-called preferred basis, in which the steady-state excitonic density matrix is diagonal. The NESS properties are examined by relating the preferred-basis description to the descriptions in the site or excitonic bases. Focusing on a model photosynthetic dimer, we find that the NESS in the limit of long trapping time is quite similar to the excited-state equilibrium, in which the stationary coherences originate from the excitation--environment entanglement. For shorter trapping times, we demonstrate how the properties of the NESS can be extracted from the time-dependent description of an incoherently driven, but unloaded dimer. This relation between stationary and time-dependent pictures is valid provided that the trapping time is longer than the decay time of dynamic coherences accessible in femtosecond spectroscopy experiments.
\end{abstract}

\maketitle

\section{Introduction}
The observation of unexpectedly long-lived oscillatory features of ultrafast spectroscopic signals measured on photosynthetic pigment--protein complexes~\cite{Nature.446.782,ProcNatlAcadSci.107.12766} has generated much excitement during the last decade. The idea that (dynamic) coherences modulating these signals may be directly relevant to natural light harvesting,~\cite{AnnuRevPhysChem.60.241,JPhysChemB.115.6227} which is triggered by stationary incoherent sunlight,~\cite{ProcNatlAcadSci.109.19575,JPhysChemLett.9.2946} has been driving intense research activities in the field.~\cite{ChemPhys.532.110663,SciAdv.6.eaaz4888} Insights from ultrafast spectroscopies are indispensable in determining the underlying Hamiltonian of the system under investigation, from which the dynamics of excitation energy transfer (EET) under any excitation condition may be inferred (as long as the excitation is sufficiently weak).~\cite{NewJPhys.12.065044,ChemPhys.439.100,JPhysChemLett.9.2946,JPhysChemLett.9.1568,ChemPhys.532.110663} A number of experimental~\cite{Nature.446.782,ProcNatlAcadSci.107.12766} and theoretical~\cite{ProcNatlAcadSci.106.17255,NatPhys.6.462} studies speculating about a positive impact of coherences and entanglement on the biological EET have been performed on the so-called unloaded systems. Such systems feature no transmission of photoinduced excitations to the reaction center (RC or load) from which the excitation energy is eventually harvested. Furthermore, the time scales addressed in these studies are generally much shorter than those representative of exciton recombination, either radiative or nonradiative.~\cite{Photosynthetic-excitons-book} Direct experimental insights into the dynamics of molecular aggregates initiated by incoherent light are limited.~\cite{JPhysChemA.117.5926} Therefore, at present, the possible relevance of some sort of quantum coherence for EET under natural conditions is best examined within appropriate theoretical models.

Such models should contain a realistic description of photoexcitation by natural incoherent light, whose intensity is essentially constant from the molecular viewpoint. Thus, the physically plausible description of natural light harvesting should feature continuous generation of electronic excitations by light, their continuous delivery to the RC, and their continuous loss by recombination.~\cite{JPhysChemLett.4.362,JChemPhys.148.124114,PLOSONE.8.0057041,JPhysChemLett.9.2946,JChemPhys.138.174103,JChemPhys.152.154101} The EET is then studied from the standpoint of nonequilibrium steady states (NESSs), which arise as a result of excitation photogeneration, phonon-induced relaxation, dephasing, trapping at the RC, and recombination. Furthermore, the coupling between the radiation and absorbing pigments is, in general, weak, so that its second-order treatment is reasonable. Then, the only information we need about the radiation is its first-order correlation function,~\cite{NewJPhys.12.065044} which can be either modeled by appropriate expressions~\cite{NewJPhys.12.065044,JChemPhys.140.074104} or obtained by a suitable ensemble average.~\cite{JChemPhys.140.074104,JChemPhys.144.044103} Excitation and deexcitation events can be treated within the Born--Markov approximation~\cite{Breuer-Petruccione-book} by Lindblad dissipators,~\cite{JPhysChemLett.3.3136,JPhysB.51.054002} or by establishing the Bloch--Redfield quantum master equation.~\cite{JChemPhys.142.104107,JChemPhys.148.124114} Approaching the problem from the perspective of open quantum systems, one can introduce an appropriate spectral density of light--matter coupling,~\cite{PhysRevA.87.022106,JPhysB.50.184003,Olsina:2014,JChemPhys.152.154101} and possibly treat it even beyond the second order.~\cite{Olsina:2014}

Reasonable models of photosynthetic EET should not overlook the non-Markovian interplay between photoinduced electronic excitations and nuclear reorganization processes,~\cite{AnnuRevCondensMatterPhys.3.333,AnnuRevPhysChem.66.69} whose relevance is emphasized by ultrafast spectroscopic studies. To that end, a number of studies attempt to combine an explicit treatment of the photoexcitation step with a nonperturbative approach to the excitation--environment coupling.~\cite{JPhysChemLett.3.3136,JPhysB.51.054002,PhysRevLett.123.093601,JChemPhys.151.034114,technical-vj-tm,Olsina:2014} The method of choice for an exhaustive treatment of excitation--environment coupling are hierarchical equations of motion (HEOM).~\cite{JChemPhys.130.234111,JChemPhys.153.020901} In the accompanying paper,~\cite{technical-vj-tm} we combine HEOM with a second-order treatment of light--matter coupling for light of arbitrary properties. Our method correctly captures light-induced reorganization processes and nonequilibrium evolution of the bath between the two interactions with light.

Recently, a number of groups have suggested that stationary coherences in the energy basis (interexciton coherences) or local basis (intersite coherences) under incoherent illumination may improve the light-harvesting efficiency in photosynthetic systems.~\cite{JChemPhys.148.124114,JPhysChemLett.11.2348,NESS-Cao-20,jung2020energy} However, the majority of existing theoretical approaches to NESSs in photosynthetic light harvesting typically feature a simplified treatment of the photoexcitation~\cite{JChemPhys.138.174103,JPhysChemB.118.10588,PLOSONE.8.0057041} or a simplified treatment of excitation relaxation and dephasing.~\cite{JPhysChemB.118.10588,JChemPhys.148.124114,PLOSONE.8.0057041,NESS-Cao-20,jung2020energy} Also, efficient algorithms that avoid the explicit temporal propagation in the computation of NESSs induced by natural incoherent light have just begun to be developed.~\cite{JChemPhys.149.114104} Therefore, there is still an urge to construct theoretical methods that circumvent the disparity between the time scales of EET dynamics and incoherent excitation sources, and yet meet the two requirements outlined in the above text.

Another pertaining issue is the origin of stationary coherences, i.e., whether they are primarily induced by incoherent radiation or by the coupling to the protein environment. The authors of Ref.~\onlinecite{Olsina:2014} argued that the NESS coherences ultimately stem from the entanglement of electronic excitations with the environment. This entanglement has been systematically studied both analytically~\cite{PhysRevE.86.021109,JChemPhys.136.204120} and numerically~\cite{PhysRevE.86.021109,PhysRevLett.112.110401} within the undriven and unloaded spin--boson model. The two-level system displays noncanonical equilibrium statistics,~\cite{JChemPhys.131.225101,PhysRevB.85.115412} whose deviation from the canonical equilibrium statistics can be conveniently measured by a single parameter.~\cite{PhysRevE.86.021109} This parameter can be interpreted as the angle by which the basis in which the system's Hamiltonian (or the system--bath interaction Hamiltonian) is diagonal should be rotated to obtain the diagonal reduced density matrix (RDM). The basis in which the RDM is diagonal is thus singled out by the environment, and the corresponding basis states are known as the preferred (or pointer) states within the framework of the decoherence theory.~\cite{RevModPhys.75.715,Schlosshauer-book} The concept of preferred basis is useful whenever representation-dependent issues, such as the ones we are after in this study, arise.

In this paper, we extend the ideas developed in Ref.~\onlinecite{PhysRevE.86.021109} to examine the properties of the NESS that arises in an incoherently driven and loaded excitonic aggregate. Adapting the algorithm presented in Ref.~\onlinecite{JChemPhys.147.044105}, we devise a procedure to find the NESS of our recently proposed HEOM that incorporates incoherent photoexcitation,~\cite{technical-vj-tm} and to properly define light-harvesting efficiency under incoherent illumination. Our theoretical approach fully respects the continuity equation for excitation fluxes. The NESS is most conveniently described in the so-called preferred basis, in which the steady-state RDM is diagonal. Such a description of a driven and loaded aggregate may be regarded as analogous to the normal-mode description of a system of coupled harmonic oscillators. While our theoretical method is quite general and applicable to arbitrary excitonic networks, we investigate the properties of the NESS using the appropriately parameterized model dimer. For realistic values of the load extraction time, we conclude that light-induced coherences are completely irrelevant in the NESS, which is then close to the noncanonical equilibrium of the undriven and unloaded dimer. We find that the NESS of the driven and loaded dimer is intimately related to the dynamics of the driven, but unloaded dimer, that takes place on the time scale of the excitation trapping at the load. This close connection between the dynamic and stationary picture is correct provided that the load extraction time is longer than the time scale of coherence dephasing, which is in principle accessible in ultrafast spectroscopies.

The paper is structured as follows: The model and method are presented in Secs.~\ref{Sec:model} and~\ref{Sec:Method}, respectively. In Sec.~\ref{Sec:bases}, we discuss the relation of the preferred-basis description of the NESS to more standard descriptions conducted in the site or excitonic basis. Section~\ref{Sec:numerics} presents numerical results, with an emphasis on the relation between the time-dependent and stationary picture. In Sec.~\ref{Sec:applications}, we outline how the methodology presented in this work could be applied to multichromophoric systems. Section~\ref{Sec:discuss-conclude} concludes the paper by summarizing its principal findings.

\section{Minimal Model}
\label{Sec:model}
We consider the simplest EET system, a molecular aggregate composed of two mutually coupled chromophores (a dimer, a spin--boson-like model~\cite{PhilTransRSocA.370.3620}). Although a similar model has been repeatedly used by many authors to gain insight into fundamentals of light harvesting under incoherent illumination,~\cite{JChemPhys.138.174103,Olsina:2014,PhysRevLett.113.113601,JChemPhys.148.124114,PhysRevA.100.043411,NESS-Cao-20,jung2020energy} our analysis uses a rigorous theoretical approach to investigate in greater detail certain properties of the NESS that have not received enough attention so far. The dimer system can be considered as the minimal model of a photosynthetic antenna with delocalization in which one can study the effects of energy relaxation, dephasing, and (static) disorder in local transition energies (the so-called asymmetric dimer, see Sec.~\ref{SSec:parameterization}). To be specific, we speak about the model dimer, while we note that the model and the method to be presented are quite general and applicable to multichromophoric situations.

Electronic excitations of the model dimer are modeled within the Frenkel exciton model~\cite{Valkunas-Abramavicius-Mancal-book,May-Kuhn-book} and the corresponding Hamiltonian reads as
\begin{equation}
\label{Eq:Frenkel-H-single-exc}
H_M=\sum_j \varepsilon_{j} |l_j\rangle\langle l_j|+\sum_{jk} J_{jk}|l_j\rangle\langle l_k|.
\end{equation}
In Eq.~\eqref{Eq:Frenkel-H-single-exc}, $|l_j\rangle$ is the singly excited state localized on chromophore $j$, $\varepsilon_{j}$ is its vertical excitation energy, while $J_{jk}$ are resonance couplings (we take $J_{kk}\equiv 0$). We limit our discussion to the manifold of singly excited states, which is justified under the assumption that the driving by the radiation is sufficiently weak. The aggregate is in contact with the thermal bath, which represents its protein environment, and which is modeled as a collection of independent oscillators labelled by site index $j$ and mode index $\xi$
\begin{equation}
\label{Eq:H-B}
H_B=\sum_{j\xi}\hbar\omega_\xi b_{j\xi}^\dagger b_{j\xi}.
\end{equation}
The phonon creation and annihilation operators $b_{j\xi}^\dagger$ and $b_{j\xi}$ entering Eq.~\eqref{Eq:H-B} satisfy Bose commutation relations. The aggregate is driven by weak radiation and the generation of excitations is described in the dipole and rotating-wave approximations
\begin{equation}
\label{Eq:H-M-R-def-quantum}
H_{M-R}=-\boldsymbol{\mu}_{eg}\cdot\mathbf{E}^{(+)}-\boldsymbol{\mu}_{ge}\cdot\mathbf{E}^{(-)}.
\end{equation}
In Eq.~\eqref{Eq:H-M-R-def-quantum}, operators $\mathbf{E}^{(\pm)}$ are the positive- and negative-frequency part of the (time-independent) operator of the (transversal) electric field, while the $eg$ part of the dipole-moment operator reads as
\begin{equation}
\label{Eq:def-vec-mu-eg}
\boldsymbol{\mu}_{eg}=\boldsymbol{\mu}_{ge}^\dagger=\sum_j\mathbf{d}_{j}\:|l_j\rangle\langle g_{j}|.
\end{equation}
We assume that transition dipole moment $\mathbf{d}_{j}$ of chromophore $j$ does not depend on nuclear coordinates (Condon approximation). The interaction of photoinduced excitations with the environment is taken to be in Holstein form, i.e., it is local and linear in oscillator displacements
\begin{equation}
\label{Eq:H-M-B}
\begin{split}
H_{M-B}&=\sum_j |l_j\rangle\langle l_j| \sum_\xi g_{j\xi}\left(b_{j\xi}^\dagger+b_{j\xi}\right)\equiv\sum_j V_j u_j.
\end{split}
\end{equation}

We assume there are two possible channels through which photogenerated excitons may decay. The first one is their transfer to the
charge-separated state in the
RC, in which case they are usefully harvested. On the other hand, exciton recombination, either radiative or nonradiative, is detrimental to the efficiency of EET. While our description of exciton photogeneration and the subsequent phonon-induced relaxation is exact, see Sec.~\ref{Sec:Method} and Ref.~\onlinecite{technical-vj-tm}, the description of exciton trapping by the RC and exciton recombination is only effective and relies on the results of more elaborate treatments performed in Refs.~\onlinecite{JChemTheorComput.7.2166,NewJPhys.20.113040}. There, it is realized that EET from one chromophore to another, as well as the radiative decay to the ground state, are actually mediated by the bath of environmental photons. Performing a second-order treatment of the appropriate interaction Hamiltonians, one ends up with effective Liouville superoperators $\mathcal{L}_\mathrm{RC}$ and $\mathcal{L}_\mathrm{rec}$ that describe respectively the excitation trapping and recombination on the level of reduced excitonic dynamics. In the following sections, we provide more details on the form of these effective Liouvillians and the manner in which they enter our description.

\section{Methods}
\label{Sec:Method}
We use our exact description of weak-light-induced exciton dynamics in molecular aggregates, which is developed in the accompanying paper.~\cite{technical-vj-tm} The radiation correlation function, which is the only property of the radiation entering our reduced description, is modeled by the following expression~\cite{Loudon-book}
\begin{equation}
\label{Eq:G-collision-broadended-CW}
\begin{split}
G^{(1)}(\tau)&=\left\langle E^{(-)}(\tau)E^{(+)}(0)\right\rangle_R\\
&=I_0\exp\left(\im\omega_c\tau-\tau/\tau_c\right),
\end{split}
\end{equation}
where $I_0$, $\omega_c$, and $\tau_c$ are the intensity, central frequency, and coherence time of the radiation, respectively. The radiation is assumed to have well defined directions of propagation and polarization. The weak-light assumption underlying our theoretical approach is well satisfied in a wide variety of photosynthetically relevant situations. For example, for a bacteriochlorophyll molecule irradiated by ambient sunlight, the excitation--light interaction may be estimated to be of the order of $10^{-3}\:\mathrm{cm}^{-1}$, see the accompanying paper~\cite{technical-vj-tm} and Ref.~\onlinecite{JPhysB.51.054002}. This energy scale is much smaller than the typical energy scales ($\sim 10-100\:\mathrm{cm}^{-1}$) of resonance couplings or the excitation--environment coupling. Also, the number of photons per unit time incident on a single photosynthetic complex under sunlight at the surface of the Earth can be estimated to be of the order of $1000\:\mathrm{s}^{-1}$, see the accompanying paper~\cite{technical-vj-tm} and Ref.~\onlinecite{JChemPhys.152.154101}. In other words, the corresponding time scale is orders of magnitude longer than time scales typical for excitation transport, trapping at the load, and recombination.

For the sake of simplicity, we assume that the baths on both sites are identical, but uncorrelated. The bath correlation function, which is the only property of the bath entering the reduced description, can be decomposed into the optimized exponential series~\cite{MolPhys.116.780} ($t\geq 0$)
\begin{equation}
\label{Eq:C-j-in-exp-decay}
\begin{split}
C(t)=\left\langle u_j(t) u_j(0)\right\rangle_B
=\sum_{m=0}^{K-1} c_{m}\:\e^{-\mu_{m}t}+2\Delta\delta(t).
\end{split}
\end{equation}
In Eq.~\eqref{Eq:C-j-in-exp-decay}, the collective bath coordinate $u_j$ is defined in Eq.~\eqref{Eq:H-M-B}, the expansion coefficients $c_{m}$ may be complex, while the corresponding decay rates $\mu_{m}$, as well as the white-noise-residue strength $\Delta$, are assumed to be real and positive. The bath correlation function is usually expressed in terms of the environmental spectral density $J(\omega)$ [$\beta=(k_BT)^{-1}$, where $T$ is the temperature]
\begin{equation}
\label{Eq:C-j-in-J-j}
\begin{split}
C(t)=\frac{\hbar}{\pi}\int_{0}^{+\infty}\dif\omega\:J(\omega)\:\frac{\e^{\im\omega t}}{\e^{\beta\hbar\omega}-1},
\end{split}
\end{equation}
which conveniently combines information on the density of environmental-mode states and the respective coupling strengths to electronic excitations.~\cite{May-Kuhn-book,Valkunas-Abramavicius-Mancal-book} We explicitly treat only $K=N_\mathrm{BE}+N_J$ terms in Eq.~\eqref{Eq:C-j-in-exp-decay}, where $N_\mathrm{BE}$ and $N_J$ are the numbers of explicitly treated poles of the Bose--Einstein function and the bath spectral density, respectively. We assume the environmental spectral density of the overdamped Brownian oscillator
\begin{equation}
\label{Eq:def-J}
 J(\omega)=2\lambda\frac{\omega\gamma}{\omega^2+\gamma^2},
\end{equation}
where $\lambda$ is the reorganization energy, while $\gamma^{-1}$ is the characteristic time scale for the decay of the bath correlation function $C(t)$.

The exponential decompositions embodied in Eqs.~\eqref{Eq:G-collision-broadended-CW} and~\eqref{Eq:C-j-in-exp-decay} enable us to formulate the problem as HEOM incorporating photoexcitation. As demonstrated in Ref.~\onlinecite{technical-vj-tm}, the hierarchy consists of two parts, one in the $eg$ sector, and another in the $ee$ sector. Each density matrix $\sigma_\mathbf{n}(t)$ is uniquely characterized by vector $\mathbf{n}$ of non-negative integers $n_{j,m}$, where index $j$ enumerates chromophores, while index $m$ counts terms in the decomposition of $C(t)$ [Eq.~\eqref{Eq:C-j-in-exp-decay}]. In order to describe excitation harvesting by the RC and recombination, we augment our formalism by effective Liouvillians describing these two processes. As demonstrated in Ref.~\onlinecite{JChemTheorComput.7.2166}, these Liouvillians appear on each level of HEOM. Performing the appropriate rescalings, which ensure that auxiliary density operators are all dimensionless and consistently smaller in deeper levels of the hierarchy,~\cite{JChemPhys.130.084105} we obtain the following equations describing the NESS we are interested in ($\gamma_\mathbf{n}=\sum_{j,m}n_{j,m}\mu_m$)
\begin{equation}
\label{Eq:time-evol-y-2nd-rescaling}
\begin{split}
 0&=-\frac{\im}{\hbar\gamma}\left[H_M,\sigma_{eg,\mathbf{n}}^{ss}\right]-\frac{\gamma_\mathbf{n}}{\gamma}\sigma_{eg,\mathbf{n}}^{ss}\\
 &+\left(\im\frac{\omega_{c}}{\gamma}-\left(\tau_{c}\gamma\right)^{-1}\right)\sigma_{eg,\mathbf{n}}^{ss}
 -\frac{\Delta}{\hbar^2\gamma}\sum_{j}V_j\sigma_{eg,\mathbf{n}}^{ss}\\
 &+\delta_{\mathbf{n},\mathbf{0}}\frac{\im}{\hbar\gamma}I_{0}\mu_{eg}\\
 &+\im\sum_{j}\sum_{m=0}^{K-1}\sqrt{1+n_{j,m}}\sqrt{\frac{\left|c_{m}\right|}{(\hbar\gamma)^2}} V_j\sigma_{eg,\mathbf{n}_{j,m}^+}^{ss}\\
 &+\im\sum_{j}\sum_{m=0}^{K-1}\sqrt{n_{j,m}}\frac{c_{m}/(\hbar\gamma)^2}{\sqrt{\left|c_{m}\right|/(\hbar\gamma)^2}}V_j\sigma_{eg,\mathbf{n}_{j,m}^-}^{ss},
\end{split}
\end{equation}

\begin{equation}
\label{Eq:time-evol-n-2nd-rescaling}
 \begin{split}
 0&=-\frac{\im}{\hbar\gamma}\left[H_M,\sigma_{ee,\mathbf{n}}^{ss}\right]
  -\frac{\gamma_\mathbf{n}}{\gamma}\sigma_{ee,\mathbf{n}}^{ss}\\
  &-\frac{\Delta}{\hbar^2\gamma}\sum_{j}V_j^\times V_j^\times\sigma_{ee,\mathbf{n}}^{ss}\\
  &+\frac{\im}{\hbar\gamma}\mu_{eg}\:\sigma_{eg,\mathbf{n}}^{ss\dagger}-
  \frac{\im}{\hbar\gamma}\sigma_{eg,\mathbf{n}}^{ss}\:\mu_{eg}^\dagger\\
  &+\gamma^{-1}\mathcal{L}_\mathrm{rec}[\sigma_{ee,\mathbf{n}}^{ss}]+\gamma^{-1}\mathcal{L}_\mathrm{RC}[\sigma_{ee,\mathbf{n}}^{ss}]\\
  &+\im\sum_{j}\sum_{m=0}^{K-1}\sqrt{1+n_{j,m}}\sqrt{\frac{\left|c_{m}\right|}{(\hbar\gamma)^2}}\:V_j^\times\sigma_{ee,\mathbf{n}_{j,m}^+}^{ss}\\
  &+\im\sum_{j}\sum_{m=0}^{K-1}\sqrt{n_{j,m}}\frac{c_{m}/(\hbar\gamma)^2}{\sqrt{\left|c_{m}\right|/(\hbar\gamma)^2}}\:V_j\sigma_{ee,\mathbf{n}_{j,m}^-}^{ss}\\
  &-\im\sum_j\sum_{m=0}^{K-1}\sqrt{n_{j,m}}
  \frac{c_{m}^*/(\hbar\gamma)^2}{\sqrt{\left|c_{m}\right|/(\hbar\gamma)^2}}\:\sigma_{ee,\mathbf{n}_{j,m}^-}^{ss}V_j .
 \end{split}
\end{equation}

In the NESS, the continuity equation for exciton currents should be valid, i.e., the number of generated excitons per unit time must balance the sum of the recombination and trapping exciton fluxes. This is physically clear, and it is seen from a more formal perspective by taking the trace (with respect to the electronic system of interest) of Eq.~\eqref{Eq:time-evol-n-2nd-rescaling} in which $\mathbf{n}=\mathbf{0}$. This results in
\begin{equation}
\label{Eq:continuity}
J_\mathrm{gen}-J_\mathrm{RC}-J_\mathrm{rec}=0.
\end{equation}
In the continuity equation [Eq.~\eqref{Eq:continuity}], we define all currents to be positive, while the sign is determined by the "direction" of the current ($+/-$ if it leads to an increase/a decrease in the exciton number). In more detail, the definitions of currents, which are dimensionless in our description, are
\begin{equation}
J_\mathrm{gen}=\frac{2}{\hbar\gamma}\mathrm{Im}\:\mathrm{Tr}_M\left\{\sigma_{eg,\mathbf{0}}^{ss}\mu_{eg}^\dagger\right\},
\end{equation}
\begin{equation}
\label{Eq:J_RC}
J_\mathrm{RC}=-\gamma^{-1}\mathrm{Tr}_M\left\{\mathcal{L}_\mathrm{RC}\left[\sigma_{ee,\mathbf{0}}^{ss}\right]\right\},
\end{equation}
\begin{equation}
J_\mathrm{rec}=-\gamma^{-1}\mathrm{Tr}_M\left\{\mathcal{L}_\mathrm{rec}\left[\sigma_{ee,\mathbf{0}}^{ss}\right]\right\}.
\end{equation}

Let us immediately note that currents $J_\mathrm{gen},J_\mathrm{RC},$ and $J_\mathrm{rec}$ are written in a basis-invariant manner. This feature is quite appealing, since one can express currents in terms of populations and coherences in any particular basis. The light-harvesting efficiency is defined as
\begin{equation}
\label{Eq:def-eta}
    \eta=\frac{J_\mathrm{RC}}{J_\mathrm{gen}}.
\end{equation}
As discussed in the accompanying paper~\cite{technical-vj-tm} and in Sec.~\ref{Sec:applications}, for realistic values of the light coherence time $\tau_c\sim 1\:\mathrm{fs}$, the expression for the generation current may be further simplified to
\begin{equation}
 J_\mathrm{gen}=2\frac{I_0\gamma\tau_c}{(\hbar\gamma)^2}\mathrm{Tr}_M\left\{\mu_{eg}|g\rangle\langle g|\mu_{ge}\right\}.
\end{equation}
In other words, possible enhancements in $\eta$ due to coherences in any basis ultimately originate from the expression for the trapping Liouvillian $\mathcal{L}_\mathrm{RC}$, as detailed in the following section.

\section{Different Bases}
\label{Sec:bases}
In the literature on the physics of photosynthetic light harvesting, two bases play special roles. The first one is basis $\{|l_j\rangle|j\}$ of singly excited states localized on single chromophores (local or site basis). While Hamiltonian parameters are usually known in the local basis, the description of, e.g., absorption properties of a photosynthetic aggregate is usually performed in the excitonic basis $\{|x_j\rangle|j\}$, i.e., in the basis of stationary states of the isolated-aggregate Hamiltonian $H_M$. Claims about possible impact of coherences on the efficiency of photosynthetic light harvesting are usually made with the coherences in the excitonic or local basis in mind.

Tomasi and Kassal have recently classified different types of possible coherent enhancements of light harvesting efficiency according to the basis in which the excitation decay mechanisms are defined.~\cite{JPhysChemLett.11.2348} Let us now focus on the trapping at the RC. Two forms for the Liouvillian $\mathcal{L}_\mathrm{RC}[\rho]$ are widely used in the literature.~\cite{JChemPhys.138.174103,JPhysChemB.118.10588,JChemPhys.148.124114,NESS-Cao-20,jung2020energy} If site $j_0$ is closest to the RC, so that it is essentially the sole site coupled to it, one uses the so-called localized-trapping Liouvillian~\cite{JChemPhys.138.174103,JPhysChemB.118.10588,JChemPhys.148.124114}
\begin{equation}
\label{Eq:loc-trap}
\begin{split}
    \mathcal{L}^\mathrm{loc}_\mathrm{RC}[\rho]=\tau_\mathrm{RC}^{-1}\left(|\mathrm{RC}\rangle\langle l_{j_0}|\rho|l_{j_0}\rangle\langle\mathrm{RC}|-\frac{1}{2}\left\{|l_{j_0}\rangle\langle l_{j_0}|,\rho\right\}\right),
\end{split}    
\end{equation}
where $\tau_\mathrm{RC}$ is the characteristic time scale on which the populations are delivered from site $j_0$ to the RC. However, even in such a situation, the excitation transfer to the RC should be regarded as the multichromophoric F\"orster transfer,~\cite{PhysRevLett.92.218301,JChemPhys.142.094106} so that it is more appropriate to employ the so-called delocalized-trapping Liouvillian~\cite{JPhysChemB.118.10588,JChemPhys.148.124114}
\begin{equation}
\label{Eq:deloc-trap}
\begin{split}
        \mathcal{L}^\mathrm{deloc}_\mathrm{RC}[\rho]&=\tau_\mathrm{RC}^{-1}\sum_j\left|\left\langle x_j|l_{j_0}\right\rangle\right|^2\\
        &\times\left(|\mathrm{RC}\rangle\langle x_{j}|\rho|x_{j}\rangle\langle\mathrm{RC}|-\frac{1}{2}\left\{|x_{j}\rangle\langle x_{j}|,\rho\right\}\right).
\end{split}
\end{equation}

The recombination occurs due to both radiative and nonradiative processes. In other words, the lifetime of chromophores' excited states is determined not only by the fluorescent decay, but to a great extent also by the conversion of the singlet states to triplets. While the radiative recombination is most naturally described in the excitonic basis, the nonradiative processes likely occur in a different basis of states, depending on how the triplet states are coupled among themselves. Overall, it is not possible to say that recombination occurs straightforwardly in neither the delocalized, nor local basis. For definiteness, we assume that the (nonradiative) recombination may occur from each site with the same rate constant $\tau_\mathrm{rec}$, and the appropriate Liouvillian reads as
\begin{equation}
\label{Eq:L-rec}
 \mathcal{L}_\mathrm{rec}^\mathrm{loc}[\rho]=\tau_\mathrm{rec}^{-1}\sum_j\left(|g\rangle\langle l_j|\rho|l_j\rangle\langle g|-\frac{1}{2}\left\{|l_j\rangle\langle l_j|,\rho\right\}\right).
\end{equation}
The fact that we have chosen the local version of the recombination should not, however, influence our results, because the recombination process is generally much slower than every other process, it occurs on the time scale of nanoseconds, i.e., several orders of magnitude slower than trapping and other processes, see Sec.~\ref{Sec:numerics}. In situations in which the radiative decay is the major loss channel, it may be more appropriate to consider the recombination Liouvillian in the excitonic basis
\begin{equation}
\label{Eq:L-rec-deloc}
    \mathcal{L}_\mathrm{rec}^\mathrm{deloc}[\rho]=\sum_j\tau_j^{-1}\left(|g\rangle\langle x_j|\rho|x_j\rangle\langle g|-\frac{1}{2}\left\{|x_j\rangle\langle x_j|,\rho\right\}\right),
\end{equation}
where $\tau_j^{-1}$ is the Weisskopf--Wigner spontaneous emission rate~\cite{JPhysB.51.054002} from excitonic state $|x_j\rangle$.

If we assume the trapping at the RC to be governed by Eq.~\eqref{Eq:loc-trap}, the trapping current $J_\mathrm{RC}$ in Eq.~\eqref{Eq:J_RC} depends only on site populations, and on both exciton populations and interexciton coherences. If, on the other hand, we assume that the trapping is governed by Eq.~\eqref{Eq:deloc-trap}, $J_\mathrm{RC}$ is expressed in terms of exciton populations only, while its expression in the local basis contains both site populations and intersite coherences. This has been recognized in recent studies, which ascertain that possible coherent enhancements of the efficiency can be achieved only when the coherence occurs in a basis different from that in which the trapping or recombination are modeled.~\cite{JChemPhys.148.124114,JPhysChemLett.11.2348}

However, we feel that the notion of coherent efficiency enhancements is not defined well enough. The word "enhancement" would suggest that there is a reference value of the efficiency (which should be smaller than unity), with respect to which we can expect to achieve an enhancement. Moreover, since the enhancement is supposed to be coherent, one could imagine that the aforementioned reference value should depend on populations only, so that the subsequent inclusion of coherences should enhance the efficiency above that reference value. Whatever the form of the effective trapping and recombination Liouvillians is, there will always be a basis in which $J_\mathrm{RC}$ is entirely expressed in terms of basis-state populations only. Such a basis will be denoted as $\{|p_j\rangle|j\}$ and termed the preferred basis of the NESS under investigation. The trapping current is then expressed only in terms of the RDM diagonal elements, so that the efficiency value calculated in the preferred basis could be regarded as the reference value above which coherent enhancements due to non-zero values of coherences in some other basis (e.g., in the local or excitonic basis) may be possible. However, our definition of the efficiency [Eq.~\eqref{Eq:def-eta}] is basis-independent, and the fact that the coherences in the excitonic or local basis are non-zero should not be expected to bring about any efficiency enhancements.

While the above discussion about coherent efficiency enhancements is quite general, it will not be the main topic of our numerical investigations, which focus on a model photosynthetic dimer and in which the efficiency is close to unity, see Sec.~\ref{SSec:parameterization}. Our central question is how the stationary state of the incoherently driven and loaded molecular system looks like, and how it relates to the two standard pictures, the picture of local states and the delocalized-states picture. Nevertheless, to the best of our knowledge, our work is among the first works that properly describe the principal physical features of excitation harvesting under incoherent light and properly define the light-harvesting efficiency.~\cite{JChemPhys.152.154101,NESS-Cao-20,jung2020energy} The arguments of the previous paragraph suggest that, once a proper description of the state in which the photosynthetic systems under incoherent illumination and load find themselves, the involvement of the coherences in the description of photosynthetic light harvesting should be critically reassessed. We thus believe that future applications of our NESS methodology to larger systems, see Sec.~\ref{Sec:applications}, could significantly contribute to the debate on possible coherent efficiency enhancements.

The excited-state sector of the steady-state RDM $\rho_{ee}^{ss}$ in the preferred basis reads as
\begin{equation}
    \rho_{ee}^{ss}\equiv\sigma_{ee,\mathbf{0}}^{ss}=\sum_j p_j\:|p_j\rangle\langle p_j|.
\end{equation}
The preferred basis is determined by the competition between excitation generation, pure dephasing, energy relaxation, excitation trapping at the RC and recombination.

The excitonic (or site) basis and the preferred basis of the NESS are connected through a unitary transformation. For our model dimer, the most general transformation of that kind can be parameterized by four real parameters, so that the basis vectors in the preferred basis are expressed in terms of the basis vectors in the excitonic basis as follows~\cite{Murnaghan-book} 
\begin{equation}
\begin{split}
    \begin{pmatrix}
    |p_0\rangle\\
    |p_1\rangle
    \end{pmatrix}&=e^{i\varphi_{px}/2}\begin{pmatrix}
    e^{i\psi_{px}} & 0\\
    0 & e^{-i\psi_{px}}
    \end{pmatrix}\\
    &\times\begin{pmatrix}
    \cos\theta_{px} & \sin\theta_{px}\\
    -\sin\theta_{px} & \cos\theta_{px}
    \end{pmatrix}\\
    &\times\begin{pmatrix}
    e^{i\Delta_{px}} & 0\\
    0 & e^{-i\Delta_{px}}
    \end{pmatrix}
    \begin{pmatrix}
    |x_0\rangle\\
    |x_1\rangle
    \end{pmatrix}
\end{split}
\end{equation}
Due to the phase freedom, we can immediately remove parameters $\varphi_{px}$ and $\psi_{px}$ from further discussion, so that we are left with only two parameters, $\theta_{px}$ and $\Delta_{px}$. From our subsequent discussion, it will emerge that the rotation angle $\theta_{px}$ is closely related to the analogous rotation angle which measures the deviation from the non-canonical statistics in the undriven and unloaded system (no generation, trapping, and recombination).~\cite{PhysRevE.86.021109} The phase $\Delta_{px}$ is intimately connected to the rates of excitation trapping and recombination, which remove excitations from the system.

The parameters $\theta_{px}$ and $\Delta_{px}$ can be related to the Bloch angles $\theta_B^x$ and $\phi_B^x$ that are commonly used to characterize the basis in which the RDM is diagonal.~\cite{PhysRevE.86.021109,PhysRevLett.112.110401} Namely, one can always normalize $\rho_{ee}^{ss}$ to obtain $\widetilde{\rho}_{ee}^{ss}$ whose trace is unity and whose eigenvalues will be denoted as $\widetilde{p}_j$. For the model dimer, the operator $\widetilde{\rho}_{ee}^{ss}$ can always be expressed as
\begin{equation}
\label{Eq:general-in-1-pauli}
    \widetilde{\rho}_{ee}^{ss}=\frac{1}{2}\left(\mathbb{I}+\boldsymbol{a}\cdot\boldsymbol{\sigma}\right),
\end{equation}
where $\mathbb{I}$ is the $2\times 2$ unity matrix, while $\boldsymbol{\sigma}=\{\sigma_1,\sigma_2,\sigma_3\}$ are three Pauli matrices. Let $\sigma_3$ be diagonal in the excitonic basis (i.e., $\sigma_3=|x_0\rangle\langle x_0|-|x_1\rangle\langle x_1|$) and let $\boldsymbol{a}^x$ be the corresponding vector in Eq.~\eqref{Eq:general-in-1-pauli}. Then it can be shown that the spherical angles $\theta_B^x$ and $\phi_B^x$ on the Bloch sphere are related to parameters $\theta_{px}$ and $\Delta_{px}$ as follows
\begin{equation}
\label{Eq:bloch-theta-px}
    \cos\theta_{B}^x=\frac{a_3^x}{|\boldsymbol{a}^x|}=\mathrm{sgn}(2\widetilde{p}_0-1)\cdot\cos(2\theta_{px}),
\end{equation}
\begin{equation}
\label{Eq:bloch-delta-px}
    \tan\phi_B^x=\frac{a_2^x}{a_1^x}=-\tan(2\Delta_{px}).
\end{equation}
Equations~\eqref{Eq:bloch-theta-px} and~\eqref{Eq:bloch-delta-px} relate $\theta_{px}$ and $\Delta_{px}$ to the Bloch angles $\theta_B^x$ and $\phi_B^x$ that are straightforwardly obtained from vector $\boldsymbol{a}^x$. In a similar manner, one obtains parameters $\theta_{pl}$ and $\Delta_{pl}$ of the unitary transformation that connects the preferred and the local basis. The rotation angle $\theta_{px}$ can always be chosen in the range $(0,\pi/4)$, while $\Delta_{px}\in(-\pi/4,\pi/4)$. This is discussed in greater detail in Sec.~SI of the Supplementary Material.

\section{Numerical Results}
\label{Sec:numerics}
\subsection{Model Parameterization and Numerical Implementation}\label{SSec:parameterization}
\begin{figure}
    \centering
    \includegraphics{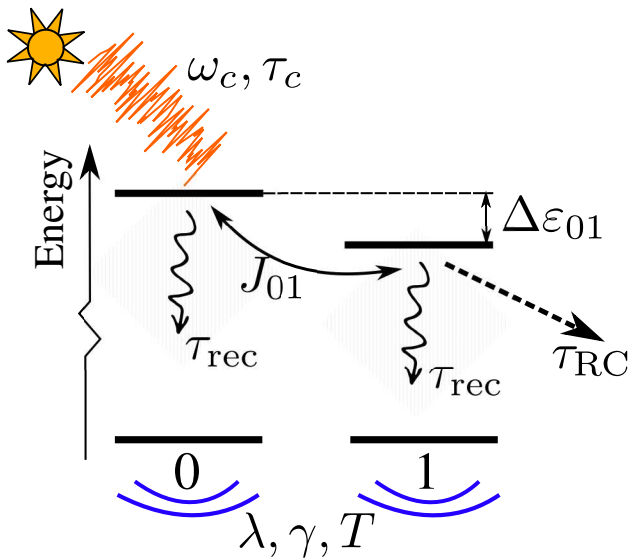}
    \caption{(Color online) Scheme of the model dimer. The electronic parameters of the dimer are the resonance coupling $J_{01}$ and the difference between the local energy levels $\Delta\varepsilon_{01}$. The dimer is excited by thermal light (schematically represented by Sun and chaotic signal) characterized by the central frequency $\omega_c$ and coherence time $\tau_c$, see Eq.~\eqref{Eq:G-collision-broadended-CW}. The transition dipole moment of site 1 is assumed to be perpendicular to the radiation polarization vector, whereas the magnitude of the projection of the transition dipole moment of site 0 onto the polarization vector is $d_{eg}$. Each chromophore is in contact with its thermal bath (schematically represented by the motion lines below chromophore numbers) characterized by the reorganization energy $\lambda$, correlation time $\gamma^{-1}$, and temperature $T$, see Eqs.~\eqref{Eq:C-j-in-J-j} and~\eqref{Eq:def-J}. The time scale of the excitation harvesting, which is governed by Eq.~\eqref{Eq:loc-trap} or Eq.~\eqref{Eq:deloc-trap}, is $\tau_\mathrm{RC}$. The time scale of the excitation loss in recombination events, which is governed by Eq.~\eqref{Eq:L-rec}, is $\tau_\mathrm{rec}$.}
    \label{fig:fig0}
\end{figure}

Numerical computations are performed on an asymmetric dimer, which is schematically presented in Fig.~\ref{fig:fig0}. The parameters of our model are summarized in Table~\ref{Tab:model_params_gen}.

\begin{table}
 \caption{Values of Model Parameters Used in Computations.}
 \label{Tab:model_params_gen}
 \centering
 \begin{tabular}{c | c}
 \hline
  $\Delta\varepsilon_{01}$ (cm$^{-1}$) & 100\\
  $J_{01}$ (cm$^{-1}$) & 100\\
  $\gamma^{-1}$ (fs) & 100\\
  $T$ (K) & 300\\
  $\hbar\omega_c$ & $\varepsilon_0$\\
  $\tau_c$ (fs) & 1.3\\
  $\tau_\mathrm{rec}$ (ns) & 1.0\\
  \hline
 \end{tabular}
\end{table}

In brief, we selectively and resonantly ($\hbar\omega_c=\varepsilon_0$) excite site 0 by weak-intensity radiation whose coherence time $\tau_c$ assumes the value representative of the natural Sunlight.~\cite{ProcPhysSoc.80.1273,IlNuovoCimento.28.401} Such a selective excitation of one site was also a feature of previous studies on model dimers.~\cite{JChemPhys.138.174103,JChemPhys.148.124114} As mentioned in Sec.~\ref{Sec:Method}, the propagation direction and the polarization vector of the radiation are assumed to be well defined. While the transition dipole moment of site 1 is assumed to be orthogonal to the polarization vector, the magnitude of the projection of the transition dipole moment of site 0 onto the polarization vector will be further denoted as $d_{eg}$. While we assume the selective excitation of site 0, the excitation trapping at the load is modeled by Eq.~\eqref{Eq:loc-trap} or Eq.~\eqref{Eq:deloc-trap} in which we assume that the load is coupled only to site 1, i.e., $j_0=1$ in Eqs.~\eqref{Eq:loc-trap} and~\eqref{Eq:deloc-trap}. Such an assumption is motivated by our wish to include the spatial transfer of excitations within our dimer model. The difference $\Delta\varepsilon_{01}=\varepsilon_0-\varepsilon_1$ between the local energy levels, resonance coupling $J_{01}$, and bath relaxation time $\gamma^{-1}$ assume values typical of the FMO complex.~\cite{RevModPhys.90.035003,JPhysChemLett.2.3045} The value of the recombination time constant $\tau_\mathrm{rec}$ is chosen on the basis of the measured exciton lifetime in the FMO complex,~\cite{PhotosynthRes.41.89,JPhysChemB.121.4700} and is similar to the value used in previous theoretical studies.~\cite{JChemPhys.148.124114,NewJPhys.10.113019,JPhysChemB.118.10588,JChemPhys.138.174103} Let us note that the recombination time scale is significantly longer than all other time scales in the problem. Also, in our model dimer, it is quite unlikely that an excitation would be prevented from reaching the RC during its lifetime. Therefore, the recombination is not probable, the precise value of $\tau_\mathrm{rec}$ is not important, and the light-harvesting efficiency will always be close to 1 in the model dimer.~\cite{JChemPhys.138.174103,NESS-Cao-20} The recombination is explicitly treated in order to formulate the continuity equation [Eq.~\eqref{Eq:continuity}], and the efficiency $\eta$ is not of primary interest in this work, which will be focused on other relevant properties of the NESS.
Within our simplified model, there is a certain level of arbitrariness in the choice of the appropriate value of $\tau_\mathrm{RC}$. Namely, a more elaborate treatment of excitation harvesting should explicitly consider both the forward and backward excitation transfer from the absorbing aggregate to the state of primary electron donor in the RC, as well as the primary charge separation, after which the excitation may be considered as usefully harvested.~\cite{Blankenship-book,IntJQuantChem.77.139,JPhysChemLett.8.6015} Previous theoretical works employing a simplified description of excitation harvesting, as we do here, typically assumed that $\tau_\mathrm{RC}$ is of the order of picoseconds.~\cite{NewJPhys.10.113019,JPhysChemLett.2.3045} On the other hand, experiments on various species of photosynthetic bacteria, as well as computational studies, suggest that the appropriate value of our parameter $\tau_\mathrm{RC}$ may be as large as a couple of tens of picoseconds.~\cite{ChemPhys.275.283,IntJQuantChem.77.139,JChemPhys.137.065101,BiophysJ.91.2778} The reported values of the reorganization energy in the FMO complex range from tens to hundreds of inverse centimeters.~\cite{BiophysJ.91.2778,PhysChemChemPhys.17.25629,PhysChemChemPhys.20.3310} Having all these things considered, the values of $\tau_\mathrm{RC}$ and $\lambda$ will be varied in wide, yet physically relevant ranges, in order to examine how they impact the properties of the NESS.

The NESS is obtained by solving coupled Eqs.~\eqref{Eq:time-evol-y-2nd-rescaling} and~\eqref{Eq:time-evol-n-2nd-rescaling} by adapting the algorithm that was introduced in Ref.~\onlinecite{JChemPhys.147.044105}. The computational approach of Ref.~\onlinecite{JChemPhys.147.044105} was developed to compute the equilibrium RDM of an undriven and unloaded system and it relies on the Jacobi iterative procedure to solve a diagonally dominant system of linear algebraic equations. The procedure is repeated until a convergence criterion, which in Ref.~\onlinecite{JChemPhys.147.044105} was related to the magnitude of the ADO elements, is satisfied. Here, we deal with a driven and loaded system, and our computations are terminated once the continuity equation [Eq.~\eqref{Eq:continuity}] is satisfied with a desired numerical accuracy. More details on our numerical scheme to compute the NESS of a driven and loaded dimer can be found in Sec.~SII of the Supplemental Material.

\subsection{Results: Long Trapping Times}\label{SSec:long-trapping-times}

\begin{figure}
    \centering
    \includegraphics[scale=1.05]{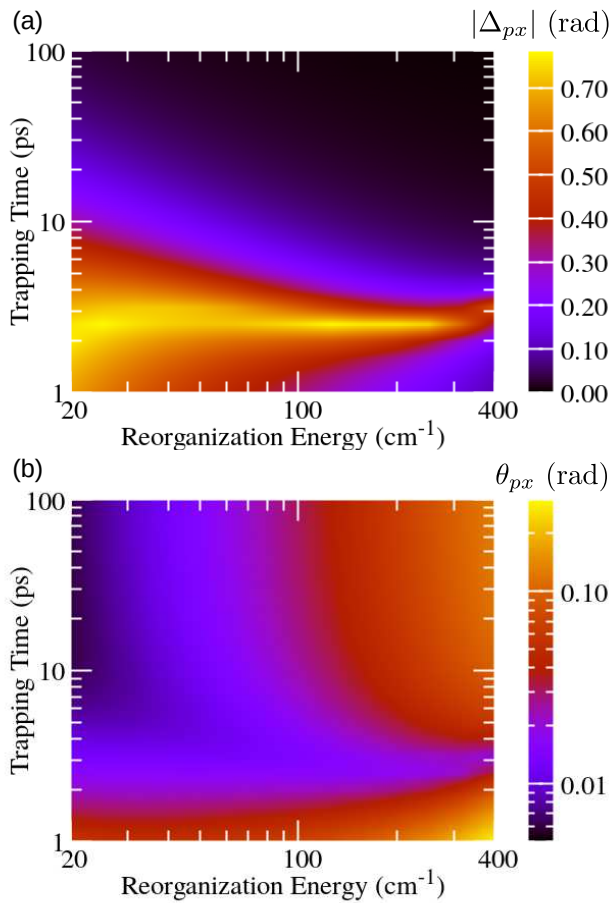}
    \caption{(Color online) Dependence of transformation parameter (a) $\Delta_{px}$ and (b) $\theta_{px}$ between the preferred and excitonic basis on the reorganization energy $\lambda$ and the trapping time $\tau_\mathrm{RC}$ at the RC. Trapping at the RC is governed by the localized-trapping Liouvillian [Eq.~\eqref{Eq:loc-trap}], while the recombination is described by the Liouvillian in Eq.~\eqref{Eq:L-rec}. Both axes feature logarithmic scale, the scale of the color bar in (a) is linear, while that of the color bar in (b) is logarithmic. The maximal value on the color bar in (a) is $\pi/4$.}
    \label{fig:fig1}
\end{figure}

\begin{figure}
    \centering
    \includegraphics[scale=1.05]{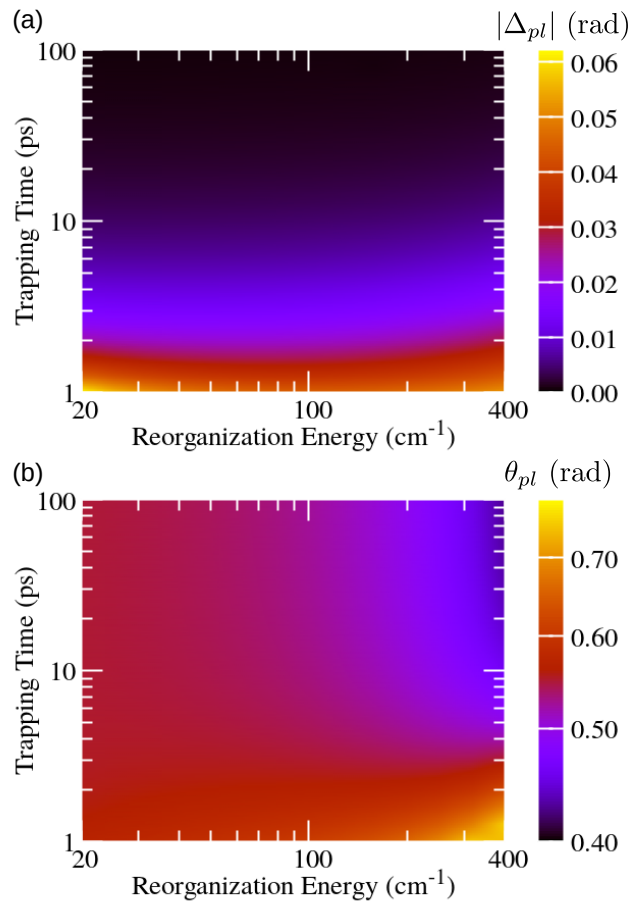}
    \caption{(Color online) Dependence of transformation parameter (a) $\Delta_{pl}$ and (b) $\theta_{pl}$ between the preferred and local basis on the reorganization energy $\lambda$ and the trapping time $\tau_\mathrm{RC}$ at the RC. Trapping at the RC is governed by the localized-trapping Liouvillian [Eq.~\eqref{Eq:loc-trap}], while the recombination is described by the Liouvillian in Eq.~\eqref{Eq:L-rec}. Both axes feature logarithmic scale, the scale of the color bar in (a) is linear, while that of the color bar in (b) is logarithmic. The maximal value on the color bar in (b) is $\pi/4$.}
    \label{fig:fig2}
\end{figure}

\begin{figure}
    \centering
    \includegraphics[scale=0.9]{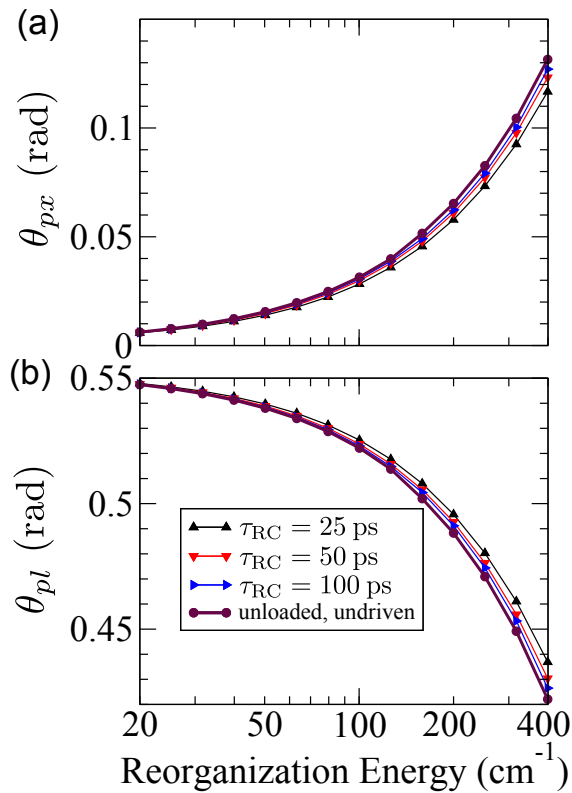}
    \caption{(Color online) Dependence of transformation parameter (a) $\theta_{px}$ (between the preferred and excitonic basis) and (b) $\theta_{pl}$ (between the preferred and site basis) on the reorganization energy $\lambda$ for fixed values of the trapping time at the RC ($\tau_\mathrm{RC}=25,\:50$ and 100 ps) and for the undriven and unloaded dimer (formally, $\tau_\mathrm{RC},\tau_\mathrm{rec}\to+\infty$). Trapping at the RC is governed by the localized-trapping Liouvillian [Eq.~\eqref{Eq:loc-trap}], while the recombination is described by the Liouvillian in Eq.~\eqref{Eq:L-rec}.}
    \label{fig:fig3}
\end{figure}

Figure~\ref{fig:fig1} (Fig.~\ref{fig:fig2}) summarizes the dependence of the parameters of the unitary transformation between the preferred and excitonic (local) basis on the reorganization energy and the trapping time at the RC. The trapping is assumed to be governed by the localized Liouvillian, see Eq.~\eqref{Eq:loc-trap}, while the recombination is described by the Liouvillian in Eq.~\eqref{Eq:L-rec}. When the trapping at the RC is so slow that $\tau_\mathrm{RC}$ is (much) longer than characteristic time scales for excitation dephasing and energy relaxation, phases $\Delta_{px}$ and $\Delta_{pl}$ tend to zero, see Figs.~\ref{fig:fig1}(a) and~\ref{fig:fig2}(a) for $\tau_\mathrm{RC}\sim 20-100$ ps. These time scales are still much shorter than those relevant for recombination. Therefore, the obtained NESS is expected to be quite similar to the equilibrium state of an undriven and unloaded aggregate.~\cite{Olsina:2014} To confirm this expectation, in Figs.~\ref{fig:fig3}(a) and~\ref{fig:fig3}(b) we plot angles $\theta_{px}$ and $\theta_{pl}$ as functions of the reorganization energy for different values of $\tau_\mathrm{RC}$. We conclude that, as $\tau_\mathrm{RC}$ is increased, angles $\theta_{px}$ and $\theta_{pl}$ tend to the values specific of the thermal equilibrium of undriven and unloaded dimer (in which we may formally identify $\tau_\mathrm{RC},\tau_\mathrm{rec}\to+\infty$). For small reorganization energies, angle $\theta_{px}$ tends to zero, see Fig.~\ref{fig:fig3}(a), and the preferred basis is close to the excitonic basis. At the same time, the limiting value reached by $\theta_{pl}$ as the reorganization energy is decreased, see Fig.~\ref{fig:fig3}(b), corresponds to the angle of the rotation by which the excitonic basis is transformed into the local basis
(the mixing angle $\theta_{xl}$ is given as $\tan(2\theta_{xl})=2J_{01}/\Delta\varepsilon_{01}$).
As the reorganization energy is increased, the preferred basis continuously changes from the excitonic basis [in which $H_M$ is diagonal, see Eq.~\eqref{Eq:Frenkel-H-single-exc}] towards the local basis [in which $H_{M-B}$ is diagonal, see Eq.~\eqref{Eq:H-M-B}].~\cite{PhysRevE.86.021109} Therefore, the magnitude of $\theta_{px}$ increases, see Fig.~\ref{fig:fig3}(a), while the magnitude of $\theta_{pl}$ decreases, see Fig.~\ref{fig:fig3}(b), with increasing reorganization energy.

\subsection{Results: Short Trapping Times. Relation between Stationary and Time-Dependent Pictures}\label{SSec:short-trapping-times}

On the other hand, when the trapping at the RC is faster, phases $\Delta_{px}$ and $\Delta_{pl}$ increase in magnitude, see Figs.~\ref{fig:fig1}(a) and~\ref{fig:fig2}(a), while the values of angles $\theta_{px}$ and $\theta_{pl}$ start to deviate from the respective values in the thermal equilibrium, see Figs.~\ref{fig:fig1}(b) and~\ref{fig:fig2}(b). These deviations are more pronounced as the trapping time is decreased and the reorganization energy is increased, see the lower right corners of Figs.~\ref{fig:fig1}(b) and~\ref{fig:fig2}(b). At the same time, the magnitude of phase $\Delta_{px}$ is large in the region of fast trapping and relatively small reorganization energy, see the lower left corner of Fig.~\ref{fig:fig1}(a), while the increase that $|\Delta_{pl}|$ displays as the trapping time is reduced is virtually the same for all considered values of reorganization energy, see Fig.~\ref{fig:fig2}(a). It was suggested that the trapping time practically determines the temporal frame in which the intrinsic dimer's dynamics is interrogated.~\cite{Olsina:2014} It is therefore interesting to examine if the dependence of the transformation parameters between the excitonic (or site) and the preferred basis of the NESS on the trapping time (vertical cuts in Figs.~\ref{fig:fig1} and~\ref{fig:fig2}) can be somehow recovered from the dynamics of the unloaded dimer initiated by suddenly turned-on incoherent light.

To gain some intuition on the relation between the stationary and time-dependent pictures, let us consider the differential equation
\begin{equation}
\label{Eq:simple-model}
\frac{df(t)}{dt}=G-Df(t).    
\end{equation}
This equation models the system, characterized by quantity $f$, that is subjected to continuous pumping (source term $G$) and continuous decay (decay rate $D$). It resembles the equation for the total excited-state population that may be obtained by taking the trace of the temporal counterpart of Eq.~\eqref{Eq:time-evol-n-2nd-rescaling} for $\mathbf{n}=\mathbf{0}$ (on the level of the RDM). The stationary point of Eq.~\eqref{Eq:simple-model} is $f^{ss}=G/D$. If one considers the driven system in the absence of decay channels, its dynamics is governed by $\frac{df^\mathrm{ul}(t)}{dt}=G$ (the superscript "ul" specifies that the system is unloaded). To solve the last equation, we assume that the driving is suddenly turned on at instant $t=0$, and that the initial condition is $f^\mathrm{ul}(0)=0$ (which would correspond to the initially unexcited system). One immediately realizes that $f^\mathrm{ul}(D^{-1})=f^{ss}$, i.e., the steady-state solution $f^{ss}$ under constant driving and decay can be obtained from the temporal evolution $f^\mathrm{ul}(t)$ of the driven system without decay channels at instant $t=D^{-1}$ corresponding to the characteristic decay time.

In the following, we develop the above-described simple argument in the situation of our interest. While the argument is quite formal, some of its parts will be specifically developed for the case of our model dimer, and in Sec.~\ref{Sec:applications} we discuss its validity in multichromophoric situations.

Let us first note that there is a hierarchy of temporal scales characteristic for the dimer’s dynamics under weak incoherent light. The shortest time scales (roughly speaking, tens to hundreds of femtoseconds) stem from the intrinsic dimer’s dynamics. The time scale of the trapping at the load is of the order of 1--10 ps in the photosynthetically relevant range of parameters. The longest time scale characterizes the excitation loss by recombination, and it is of the order of nanoseconds. The assumption of weak light actually means that the radiation is so weak that the number of incident photons per unit time is very small and the corresponding time scale is the longest time scale in the problem (we may loosely say that the light intensity is so low that the absorption of incident radiation is the rate-limiting process).

We now concentrate on the NESS Eqs.~\eqref{Eq:time-evol-y-2nd-rescaling} and~\eqref{Eq:time-evol-n-2nd-rescaling}. Within our second-order treatment of the light-matter interaction, one can first solve Eq.~\eqref{Eq:time-evol-y-2nd-rescaling} and obtain the steady-state in the $eg$ sector, $\{\sigma_{eg,\mathbf{n}}^{ss}|\mathbf{n}\}$, and then use this solution to compute the source-term in the $ee$ sector [the third term on the RHS of Eq.~\eqref{Eq:time-evol-n-2nd-rescaling}]. Given the known source term in the $ee$ sector, which will be denoted as $\mathbf{G}$, Eq.~\eqref{Eq:time-evol-n-2nd-rescaling} can be recast as
\begin{equation}
\label{Eq:ss-1}
    0=\mathbf{G}-(\hat{A}+\hat{D})\boldsymbol{\rho}^{ss},
\end{equation}
where $\boldsymbol{\rho}^{ss}$ is the HEOM-space representation of the RDM and ADMs, $\hat{A}$ is the HEOM-space representation of the hierarchical links between DMs [it comprises terms 1,~2,~5--7 on the RHS of Eq.~\eqref{Eq:time-evol-n-2nd-rescaling}], while $\hat{D}$ is the HEOM-space representation of the recombination and trapping at the load [term 4 on the RHS of Eq.~\eqref{Eq:time-evol-n-2nd-rescaling}]. Eq.~\eqref{Eq:ss-1} is solved by
\begin{equation}
\label{Eq:ss-1-sol}
  \boldsymbol{\rho}^{ss}=(\hat{A}+\hat{D})^{-1}\mathbf{G}.
\end{equation}

While the matrix $\hat{A}$ is off-diagonal in the HEOM space and non-symmetric, the matrix $\hat{D}$ is diagonal in the HEOM space. Whatever the form of the trapping and recombination Liouvillians is, all the matrix elements of $\hat{D}$ are of the same magnitude, $\sim\tau_\mathrm{RC}^{-1}$ (due to the above-mentioned hierarchy of dimer’s temporal scales, the recombination rate, $\tau_\mathrm{rec}^{-1}$, is completely irrelevant). Moreover, the eigenvalues of matrix $\hat{A}$, which represent the rates of the internal dimer’s dynamics, are much larger than $\tau_\mathrm{RC}^{-1}$. Therefore, computing the inverse $(\hat{A}+\hat{D})^{-1}$, we can regard $\hat{D}$ as a small isotropic correction to $\hat{A}$. In the lowest-order approximation, the spectral decomposition of $(\hat{A}+\hat{D})^{-1}$ can be formulated as
\begin{equation}
\label{Eq:spectral-a-plus-d}
  (\hat{A}+\hat{D})^{-1}\approx\sum_k(a_k+d_k)^{-1}\left|a_k^R\right\rangle\left\langle a_k^L\right|,  
\end{equation}
where $a_k,\left|a_k^R\right\rangle,$ and $\left\langle a_k^L\right|$ are respectively the eigenvalues, right, and left eigenvectors of matrix $\hat{A}$, while $d_k$ are the elements of $\hat{D}$ (they are all approximately the same).

Let us now focus on the temporal counterparts of Eqs.~\eqref{Eq:time-evol-y-2nd-rescaling} and~\eqref{Eq:time-evol-n-2nd-rescaling} in which the trapping and recombination Liouvillians are omitted. Within our second-order treatment of the light-matter interaction, one can first solve the dynamics in the $eg$ sector [Eq.~\eqref{Eq:time-evol-y-2nd-rescaling}], and then use this solution as a known time-dependent source term in Eq.~\eqref{Eq:time-evol-n-2nd-rescaling}, which describes the $ee$ sector we are primarily interested in. However, as advocated in the accompanying paper,~\cite{technical-vj-tm} for realistic values of the light coherence time (of the order of 1 fs for natural sunlight), this source term is approximately time-independent (see also the expression for the generation current in the limit of short coherence time of light) and thus equal to $\mathbf{G}$ introduced in the above discussion. The temporal counterpart of Eq.~\eqref{Eq:time-evol-n-2nd-rescaling} without trapping and recombination then reads as
\begin{equation}
  \partial_t\boldsymbol{\rho}^\mathrm{ul}(t)=\mathbf{G}-\hat{A}\boldsymbol{\rho}^\mathrm{ul}(t).  
\end{equation}
Assuming that the light is abruptly turned on at $t=0$ and that $\boldsymbol{\rho}^\mathrm{ul}(0)=\mathbf{0}$, the solution is
\begin{equation}
\label{Eq:time-dependent-solution}
  \boldsymbol{\rho}^\mathrm{ul}(t)=\int_0^t ds\:e^{-\hat{A}(t-s)}\:\mathbf{G}.  
\end{equation}
In the spectral representation of $\hat{A}$, the solution reads as
\begin{equation}
\label{Eq:time-dependent-solution-1}
  \boldsymbol{\rho}^\mathrm{ul}(t)=\sum_k\left(\int_0^t ds\:e^{-a_k(t-s)}\right)
\left|a_k^R\right\rangle\left\langle a_k^L\middle|\mathbf{G}\right\rangle.  
\end{equation}
It is known from the literature~\cite{JChemPhys.150.184109} that, for the spin--boson model whose coupling to the environment has the overdamped Brownian oscillator spectral density [Eq.~\eqref{Eq:def-J}], at least one of the eigenvalues of matrix $\hat{A}$ is equal to zero, while non-zero eigenvalues appear in complex conjugate pairs and have positive real parts. Assuming that $a_0=0$, the time-dependent solution for a driven but unloaded system [Eq.~\eqref{Eq:time-dependent-solution}] can be recast as
\begin{equation}
\label{Eq:t-dep-1}
    \boldsymbol{\rho}^\mathrm{ul}(t)=t\left|0^R\right\rangle\left\langle 0^L\middle|\mathbf{G}\right\rangle+
\sum_{k\neq 0}\frac{1-e^{-a_k t}}{a_k}\left|a_k^R\right\rangle\left\langle a_k^L\middle|\mathbf{G}\right\rangle.
\end{equation}
The same observations enable us to recast the spectral form of the NESS solution under driving and load as
\begin{equation}
\label{Eq:ness-1}
  \boldsymbol{\rho}^{ss}\approx\frac{1}{d_0}\left|a_0^R\right\rangle\left\langle a_0^L\middle|\mathbf{G}\right\rangle+
\sum_{k\neq 0}\frac{1}{a_k}\left(1+\frac{d_k}{a_k}\right)^{-1}\left|a_k^R\right\rangle\left\langle a_k^L\middle|\mathbf{G}\right\rangle.  
\end{equation}
Remembering that $d_0\sim\tau_\mathrm{RC}^{-1}$, and that $|d_k/a_k|$ is sufficiently smaller than 1 for all $k\neq 0$, we can approximate $e^{-a_k/\tau_\mathrm{RC}}\approx 0$ in Eq.~\eqref{Eq:t-dep-1} and $(1+d_k/a_k)^{-1}\approx 1$ in Eq.~\eqref{Eq:ness-1} to finally obtain that
\begin{equation}
\label{Eq:qed}
    \boldsymbol{\rho}^\mathrm{ul}(\tau_\mathrm{RC})\approx\boldsymbol{\rho}^{ss}\approx\tau_\mathrm{RC}\left|a_0^R\right\rangle\left\langle a_0^L\middle|\mathbf{G}\right\rangle
+\sum_{k\neq 0}\frac{1}{a_k}\left|a_k^R\right\rangle\left\langle a_k^L\middle|\mathbf{G}\right\rangle
\end{equation}
We have just demonstrated that that the dimer’s NESS can be reconstructed from the time evolution of the initially unexcited, driven, but unloaded dimer at instant $t\sim\tau_\mathrm{RC}$ after a sudden turn-on of the driving. This result establishes an interesting relationship between the stationary (with load) and time-dependent (without load) pictures under incoherent driving.

While the physical relevance of the sudden turn-on of incoherent light may be questionable, see Ref.~\onlinecite{JPhysChemLett.9.2946} and references therein, this formal argument demonstrates that there is a formal connection between this seemingly unphysical setting and the NESS picture. However unphysical the sudden turn-on may be, the final-value theorem from the theory of Laplace transforms ensures that Eq.~\eqref{Eq:time-dependent-solution}, in which the load is added, i.e., $\hat{A}\to\hat{A}+\hat{D}$, can be used to obtain the NESS in Eq.~\eqref{Eq:ss-1-sol} by temporal propagation up to sufficiently long time $t\to+\infty$.

\begin{widetext}

 \begin{figure}
    \centering
    \includegraphics{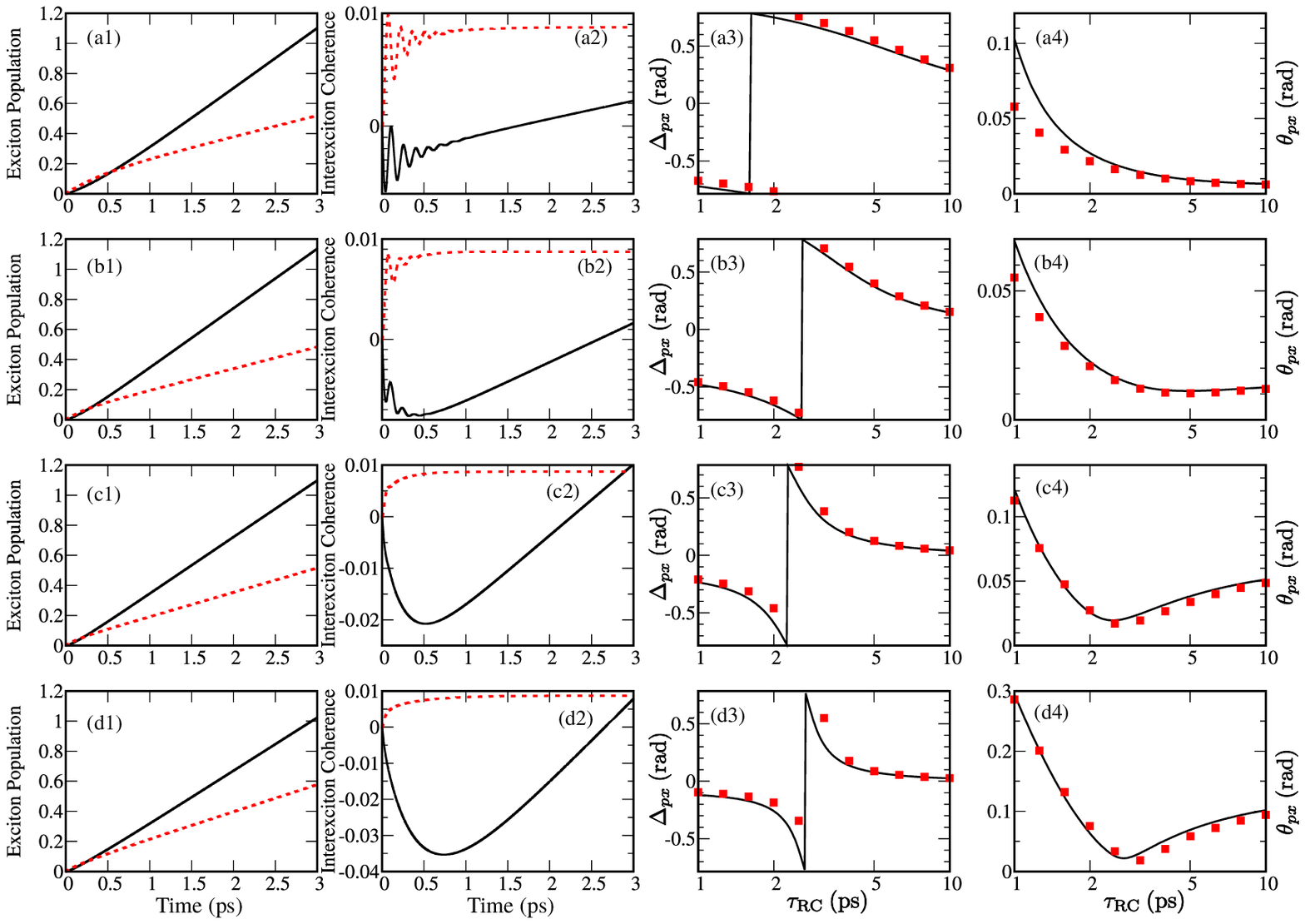}
    \caption{(Color online) (a1)--(d1): Time dependence of populations of exciton states $|x_0\rangle$ (solid line) and $|x_1\rangle$ (dashed line) of the incoherently driven and unloaded model dimer for different values of the reorganization energy. (a2)--(d2): Time dependence of the real (solid line) and imaginary (dashed line) parts of the interexciton coherence of the incoherently driven and unloaded model dimer for different values of the reorganization energy. Both exciton populations and interexciton coherences are measured in units of $I_0\:d_{eg}^2/(\hbar\gamma)^2$. The excitation is suddenly turned on at $t=0$. Dependence of the transformation parameters $\Delta_{px}$ [(a3)--(d3)] and $\theta_{px}$ [(a4)--(d4)] between the excitonic basis and the preferred basis of the NESS on the trapping time constant $\tau_\mathrm{RC}\in(1,10)\:\mathrm{ps}$ for different values of the reorganization energy. Solid lines are computed using time traces of a driven and unloaded model dimer at $t=\tau_\mathrm{RC}$, while squares emerge from the computation of the NESS using Eqs.~\eqref{Eq:time-evol-y-2nd-rescaling} and~\eqref{Eq:time-evol-n-2nd-rescaling}. The scale on the abscissa ($\tau_\mathrm{RC}$) in (a3)--(d4) is logarithmic. Trapping at the RC is governed by the localized-trapping Liouvillian [Eq.~\eqref{Eq:loc-trap}], while the recombination is described by the Liouvillian in Eq.~\eqref{Eq:L-rec}. The values of the reorganization energy are 20~cm$^{-1}$ [(a1)--(a4)], 50\:cm$^{-1}$ [(b1)--(b4)], 200\:cm$^{-1}$ [(c1)--(c4)], and 400\:cm$^{-1}$ [(d1)--(d4)].}
    \label{fig:fig4}
\end{figure}
\end{widetext}

In the following, we concentrate on a numerical demonstration of this relationship. First, we propagate the temporal counterparts of Eqs.~\eqref{Eq:time-evol-y-2nd-rescaling} and~\eqref{Eq:time-evol-n-2nd-rescaling}, in which the trapping and recombination Liouvillians are omitted, see Figs.~\ref{fig:fig4}(a1)--\ref{fig:fig4}(d2). The RDM elements [in Figs.~\ref{fig:fig4}(a1)--\ref{fig:fig4}(d2), in the excitonic basis] are measured in units of $I_0d_{eg}^2/(\hbar\gamma)^2$, and not in absolute units, as is customarily done. Our reason for choosing this unit lies in our perturbative treatment of the interaction with light, which ensures that singly excited-state populations and intraband coherences are proportional to the light intensity $I_0$, see Eq.~\eqref{Eq:G-collision-broadended-CW}, and to the excited-state oscillator strength $d_{eg}^2$, see the caption of Fig.~\ref{fig:fig0}. Analyzing Eqs.~\eqref{Eq:time-evol-y-2nd-rescaling} and~\eqref{Eq:time-evol-n-2nd-rescaling}, one can readily conclude that $I_0d_{eg}^2/(\hbar\gamma)^2$ is the natural unit in which to measure populations and coherences. The absolute value of this unit may be estimated by using the estimate for the magnitude of the excitation--light interaction given in Sec.~\ref{Sec:Method}$,\sqrt{I_0d_{eg}^2}\sim 10^{-3}\:\mathrm{cm}^{-1}$, so that for the value of $\gamma$ given in Table~\ref{Tab:model_params_gen} we obtain $I_0d_{eg}^2/(\hbar\gamma)^2\simeq 4\times 10^{-10}$. Second, we use the RDM $\sigma_{ee,\mathbf{0}}(\tau_\mathrm{RC})$ at $t=\tau_\mathrm{RC}$ to determine the transformation parameters $\Delta$ and $\theta$ by virtue of Eqs.~\eqref{Eq:general-in-1-pauli}--\eqref{Eq:bloch-delta-px}. The results emerging from these real-time computations are confronted with the results emerging from NESS computations in Figs.~\ref{fig:fig4}(a3)--\ref{fig:fig4}(d4) for a couple of values of reorganization energy. It is observed that the two methods predict quite similar values of transformation parameters $\Delta_{px}$ and $\theta_{px}$ for all the examined values of the reorganization energy and trapping time. This result, together with the RDM dynamics initiated by a sudden turn-on of incoherent radiation, can help us better understand the dependence of $\Delta_{px}$ and $\Delta_{pl}$ on $\tau_\mathrm{RC}$ for $\tau_\mathrm{RC}\sim 1-10\:\mathrm{ps}$. The discontinuous change from $-\pi/4$ to $\pi/4$ that phase $\Delta_{px}$ undergoes at around $2-3\:\mathrm{ps}$ [see the bright area in Fig.~\ref{fig:fig1}(a)] should be attributed to the fact that the real part of interexciton coherence in an incoherently driven, but unloaded dimer, becomes equal to zero at around $2-3\:\mathrm{ps}$, see solid curves in Figs.~\ref{fig:fig4}(a2)--\ref{fig:fig4}(d2). The imaginary part of the interexciton coherence saturates somewhat earlier, see dashed curves in Figs.~\ref{fig:fig4}(a2)--\ref{fig:fig4}(d2). As $\tau_\mathrm{RC}$ is further increased, $\Delta_{px}$ decreases because the real part of the interexciton coherence is increasing, see also Eq.~\eqref{Eq:bloch-delta-px}.

The relation between the NESS and RDM dynamics in real time leans on the above-mentioned hierarchy of time scales of the dimer's dynamics, i.e., on the fact that the trapping time scale is long enough. Namely, in time traces of a driven and unloaded dimer for $t\gtrsim 1\:\mathrm{ps}$, one observes that the behavior of both exciton populations [Figs.~\ref{fig:fig4}(a1)--(d1)] and interexciton coherences [Figs.~\ref{fig:fig4}(a2)--(d2)] displays certain steadiness. In other words, populations, as well as the real part of the interexciton coherence, linearly increase in time, while the imaginary part of the interexciton coherence reaches a constant value, see also the accompanying paper.~\cite{technical-vj-tm} When the trapping time constant is $\tau_\mathrm{RC}\gtrsim 1\:\mathrm{ps}$, so that at $t=\tau_\mathrm{RC}$ the steadiness has already been established, the dynamical quantities of a driven, but unloaded dimer, may be used to quite accurately reconstruct the NESS. On the other hand, when $\tau_\mathrm{RC}\lesssim 1\:\mathrm{ps}$, so that the steadiness has not been established yet, the reconstruction of the NESS from the quantities of a driven and unloaded dimer computed at $\tau_\mathrm{RC}$ would be less accurate. This is particularly clear for low values of the reorganization energy, see Figs.~\ref{fig:fig4}(a3) and~\ref{fig:fig4}(a4), when oscillations in the interexciton coherence are damped on a time scale of $\sim 200-300\:\mathrm{fs}$, see Fig.~\ref{fig:fig4}(a2). A similar situation can be expected for slow bath, when the bath correlation time $\gamma^{-1}$ is long enough.~\cite{JChemPhys.130.234111} In such cases, the reconstruction of NESS from the dynamics of the incoherently driven and unloaded dimer is not accurate because $\tau_\mathrm{RC}$ is comparable to the time scales of the intrinsic dimer's dynamics. At this point, it is useful to remember that the information extracted from ultrafast spectroscopic signals can be used to determine the Hamiltonian parameters of the system under consideration, i.e., to determine the rate constants $\mathrm{Re}\{a_k\}$ and the frequencies $\mathrm{Im}\{a_k\}$ of the oscillatory features of the intrinsic dimer's dynamics that enter Eqs.~\eqref{Eq:spectral-a-plus-d} and~\eqref{Eq:time-dependent-solution}--\eqref{Eq:qed}.~\cite{JPhysChemLett.9.2946,ChemPhys.532.110663,NewJPhys.12.065044,ChemPhys.439.100,JPhysChemLett.9.1568} The dynamics of a driven, but unloaded, system, in which the driving is abruptly turned on at $t=0$, see Eq.~\eqref{Eq:time-dependent-solution} and Figs.~\ref{fig:fig4}(a1)--\ref{fig:fig4}(d2), can be formally regarded as the interference of all possible outcomes $e^{-\hat{A}(t-s)}\mathbf{G}$ of ultrafast experiments in which the delta-like excitation is centered at instant $s\in(0,t)$ and which freely evolve for the time interval of length $t-s$. The initial condition is set by the ratios (the initial condition has to be dimensionless!) of the components of the generation vector $\mathbf{G}$, which can be shown to be basically proportional to the square of the transition dipole moments and dependent on their mutual alignments, see Sec.~\ref{Sec:applications}. The oscillatory features stemming from the abruptly turned-on incoherent light in Figs.~\ref{fig:fig4}(a2) and~\ref{fig:fig4}(b2) are thus tightly connected to the oscillatory features characteristic for the excitation by very short pulses. The time scales on which the oscillatory features can be observed thus set the lower limit on $\tau_\mathrm{RC}$ for which the above-described relationship between the stationary and dynamic pictures is valid. Therefore, the decay time of dynamical coherences observed in spectroscopies may still be relevant in the natural setting, although the dynamical coherences themselves are absent in the NESS.~\cite{JPhysChemB.118.2693} In Sec.~SIII of the Supplementary Material, we estimate the time scales characteristic of exciton decoherence by suitable fitting procedures. For the values of model parameters adopted in this work, we find that the characteristic decay times of exciton coherence are shorter than resonable values of $\tau_\mathrm{RC}$.

The previous discussion was conducted for interexciton coherences. Similar conclusions can be also reached in the site basis, see Fig.~\ref{fig:fig2}. While we have already discussed the limit of long trapping time in Fig.~\ref{fig:fig3}(b), the case of relatively short $\tau_\mathrm{RC}\sim 1-10\:\mathrm{ps}$ is analyzed in greater detail in Fig.~S2 of the Supplementary Material. The analysis is completely analogous to that accompanying Figs.~\ref{fig:fig4}(a1)--\ref{fig:fig4}(d4).

\begin{figure}
    \centering
    \includegraphics[scale=1.05]{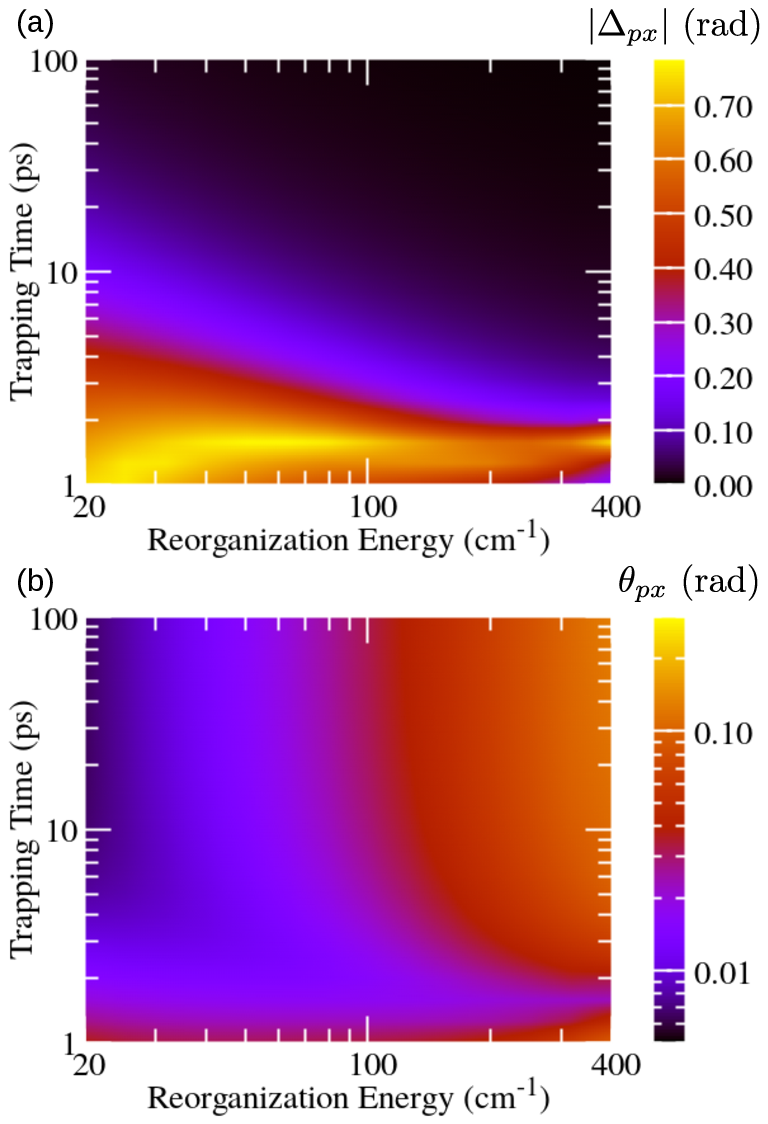}
    \caption{(Color online) Dependence of transformation parameter (a) $\Delta_{px}$ and (b) $\theta_{px}$ between the preferred and excitonic basis on the reorganization energy $\lambda$ and the trapping time $\tau_\mathrm{RC}$ at the RC. Trapping at the RC is governed by the delocalized-trapping Liouvillian [Eq.~\eqref{Eq:deloc-trap}], while the recombination is described by the Liouvillian in Eq.~\eqref{Eq:L-rec}. Both axes feature logarithmic scale, the scale of the color bar in (a) is linear, while that of the color bar in (b) is logarithmic. The maximal value on the color bar in (a) is $\pi/4$. To facilitate the comparison with Fig.~\ref{fig:fig1}, the ranges of color bars in (a) and (b) are identical to the ranges of color bars in Figs.~\ref{fig:fig1}(a) and~\ref{fig:fig1}(b), respectively.}
    \label{fig:fig6}
\end{figure}

The choice of instant $t=\tau_\mathrm{RC}$ at which time-dependent quantities are extracted to obtain the properties of the NESS is somewhat arbitrary because $\tau_\mathrm{RC}$ is not really the time, but the characteristic time scale of the trapping. This is also apparent from our formal demonstration of the relation between the stationary and time-dependent pictures, in which we only used the fact that all $d_k$ are of the order of $\tau_\mathrm{RC}^{-1}$, while their precise values were not important. Moreover, in the results presented so far, we used the trapping [Eq.~\eqref{Eq:loc-trap}] and recombination [Eq.~\eqref{Eq:L-rec}] Liouvillians that are diagonal in the local basis. It is, therefore, not obvious if and how the above-discussed relation between the dynamic and stationary pictures under incoherent driving changes when the trapping or recombination Liouvillian that is diagonal in the excitonic basis, Eqs.~\eqref{Eq:deloc-trap} and~\eqref{Eq:L-rec-deloc}, is employed. In Figs.~\ref{fig:fig6}(a) and~\ref{fig:fig6}(b), which are analogous to Figs.~\ref{fig:fig1}(a) and~\ref{fig:fig1}(b), respectively, we examine the dependence of the transformation parameters $\Delta_{px}$ and $\theta_{px}$ on $\lambda$ and $\tau_\mathrm{RC}$ under the assumption of delocalized trapping, while we retain the recombination Liouvillian in Eq.~\eqref{Eq:L-rec}. The main features of Figs.~\ref{fig:fig1}(a) and~\ref{fig:fig1}(b) are clearly recognizable in Figs.~\ref{fig:fig6}(a) and~\ref{fig:fig6}(b). This is particularly true at long trapping times. However, at short trapping times, the maximum that $|\Delta_{px}|$ reaches in Fig.~\ref{fig:fig1}(a) at $\tau_\mathrm{RC}\sim 2-3\:\mathrm{ps}$ is shifted towards $\tau_\mathrm{RC}\sim 1-2\:\mathrm{ps}$ in Fig.~\ref{fig:fig6}(a). A similar discussion applies to Fig.~\ref{fig:fig6}(b), where the decrease that $\theta_{px}$ exhibits as $\tau_\mathrm{RC}$ is increased from 1~ps is shifted to shorter trapping times with respect to Fig.~\ref{fig:fig1}(b). We believe that the maximum in $|\Delta_{px}|$, which occurs at $\tau_\mathrm{RC}\sim 1-2\:\mathrm{ps}$ for delocalized trapping, should still be interpreted to originate from the fact that the real part of the interexction coherence in the driven and unloaded dimer changes its sign on a picosecond time scale. If the real part of the interexciton coherence is equal to zero at instant $t_0$, it is not guaranteed that the magnitude of $\Delta_{px}$ (computed from the NESS) reaches it maximal value of $\pi/4$ exactly at $\tau_\mathrm{RC}=t_0$ [this is also observed in Figs.~\ref{fig:fig4}(a3)--\ref{fig:fig4}(d3)]. Our point here is that the magnitude of $\Delta_{px}$ reaches $\pi/4$ at $\tau_\mathrm{RC}\sim t_0$ irrespective of the precise form of the trapping Liouvillian.

\subsection{Further Results}\label{SSec:further-results}
In the following, we discuss how the variations in the electronic parameters of the model, in particular in the difference $\Delta\varepsilon_{01}$ between local energy levels, affect the properties of the NESS. We fix the reorganization energy to 150\:cm$^{-1}$.
\begin{figure}
    \centering
    \includegraphics[scale=1.05]{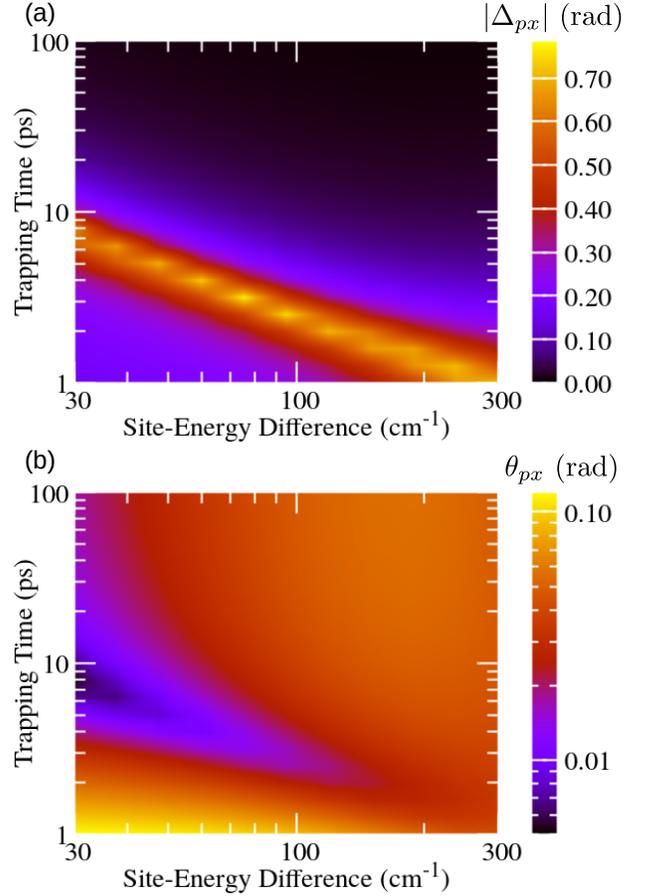}
    \caption{(Color online) Dependence of transformation parameter (a) $\Delta_{px}$ and (b) $\theta_{px}$ between the preferred and excitonic basis on the site-energy difference $\Delta\varepsilon_{01}$ and the trapping time $\tau_\mathrm{RC}$ at the RC. Trapping at the RC is governed by the localized-trapping Liouvillian [Eq.~\eqref{Eq:loc-trap}], while the recombination is described by the Liouvillian in Eq.~\eqref{Eq:L-rec}. Both axes feature logarithmic scale, the scale of the color bar in (a) is linear, while that of the color bar in (b) is logarithmic. The maximal value of the color bar in (a) is $\pi/4$. The reorganization energy assumes the value of 150\:cm$^{-1}$.}
    \label{fig:fig5}
\end{figure}
Figures~\ref{fig:fig5}(a) and~\ref{fig:fig5}(b) present the dependence of transformation parameters $\Delta_{px}$ and $\theta_{px}$ on $\tau_\mathrm{RC}$ and $\Delta\varepsilon_{01}$. We varied $\Delta\varepsilon_{01}$ from 30 to 300\:cm$^{-1}$ on the basis of the literature values of site-energy differences in the FMO complex.~\cite{BiophysJ.91.2778,RevModPhys.90.035003} Let us first focus on the long trapping times, when the magnitude of the phase $\Delta_{px}$ is small, see the upper half of Fig.~\ref{fig:fig5}(a), and the NESS obtained is quite similar to the excited-state equilibrium. For small values of $\Delta\varepsilon_{01}$, for which $J_{01}/\Delta\varepsilon_{01}\gtrsim 2$, exciton delocalization prevails over the localizing effect of the environment, which is reflected in relatively small values of the rotation angle $\theta_{px}$, see the upper left part of Fig.~\ref{fig:fig5}(b). As the local energy levels become more off-resonant, the environment-induced localization becomes more pronounced than exciton delocalization, so that the rotation angle from the excitonic basis to the preferred basis of the NESS increases. However, when $\Delta\varepsilon_{01}$ is large enough, so that $\Delta\varepsilon_{01}/J_{01}\gtrsim 2$, the excitonic basis is already localized enough and, for the chosen value of $\lambda$, the localizing effect of the environment is effectively suppressed. This leads to a decrease in the rotation angle $\theta_{px}$. As the trapping time is shortened, the deviations from the above-established picture become more pronounced. The magnitude of $\Delta_{px}$ reaches its maximum at $\tau_\mathrm{RC}=1-10\:\mathrm{ps}$ depending on the particular value of $\Delta\varepsilon_{01}$, see Fig.~\ref{fig:fig5}(a), while angle $\theta_{px}$ exhibits a minimum in the very same region of the $\Delta\varepsilon_{01}-\tau_\mathrm{RC}$ space, see Fig.~\ref{fig:fig5}(b).

\section{Possible Applications to Multichromophoric Aggregates}
\label{Sec:applications}
So far, we have employed our novel theoretical approach to study in great detail the NESS of an incoherently driven and loaded dimer. The dimer represents the only model, in which one can represent the character of the NESS by only two parameters---the angle of the basis rotation $\theta$ with respect to a chosen basis, and the phase $\Delta$ that is closely connected with the excitation decay rates---and the difference between different NESSs is easy to understand. In other words, the dimer is the only model, where relatively simple “understanding” of the role of different parameters in establishing the preferred basis can be derived. Whether our results can be translated to larger systems is, of course, an important question. This section aims at presenting the basic steps that have to be taken in order to apply our NESS formalism to multichromophoric situations.

We start from the fact that the realistic coherence time of light $\tau_c\sim 1\:\mathrm{fs}$ is much shorter than any other time scale in the problem. All our results for the dimer are obtained by solving Eqs.~\eqref{Eq:time-evol-y-2nd-rescaling} and~\eqref{Eq:time-evol-n-2nd-rescaling} and do not lean on any assumption about the coherence time $\tau_c$ entering Eq.~\eqref{Eq:G-collision-broadended-CW}. In the accompanying paper,~\cite{technical-vj-tm} we have argued that, in the limit of small $\tau_c$, the first-order light correlation function defined in Eq.~\eqref{Eq:G-collision-broadended-CW} may be replaced by
\begin{equation}
\label{Eq:G-tau-incoherent}
    G^{(1)}(\tau)=2I_0\tau_c\delta(\tau),
\end{equation}
which represents the so-called white-noise model of the radiation.~\cite{Olsina:2014} Equation~\eqref{Eq:time-evol-y-2nd-rescaling} is then omitted from further discussion, while the source term (the third term) of Eq.~\eqref{Eq:time-evol-n-2nd-rescaling} reads as~\cite{technical-vj-tm}
\begin{equation}
\label{Eq:S_WNM}
    S_\mathbf{n}^\mathrm{WNM}=\delta_{\mathbf{n},\mathbf{0}}\frac{2I_0\gamma\tau_c}{(\hbar\gamma)^2}\mu_{eg}|g\rangle\langle g|\mu_{eg}^\dagger
\end{equation}
We recall that the operator $\mu_{eg}=\mathbf{e}\cdot\boldsymbol{\mu}_{eg}$ is the projection of the dipole-moment operator $\boldsymbol{\mu}_{eg}$ onto the radiation polarization vector $\mathbf{e}$. Using the definition of $\boldsymbol{\mu}_{eg}$ in Eq.~\eqref{Eq:def-vec-mu-eg}, the matrix elements of the source term in the basis $\{|b_j\rangle|j\}$ of the single-excitation manifold can be expressed as
\begin{equation}
\label{Eq:S_WNM_matrix_elements}
\begin{split}
    \left\langle b_k\left|S_\mathbf{n}^\mathrm{WNM}\right|b_j\right\rangle=&\delta_{\mathbf{n},\mathbf{0}}\frac{2I_0\gamma\tau_c}{(\hbar\gamma)^2}\sum_{k'j'}\left(\mathbf{d}_{k'}\cdot\mathbf{e}\right)\left(\mathbf{d}_{j'}\cdot\mathbf{e}\right)\\&\times\langle b_k|l_{k'}\rangle\langle l_{j'}|b_j\rangle.
\end{split}
\end{equation}
The basis $\{|b_j\rangle|j\}$ is in principle arbitrary; it can be the local basis ($b=l$), the excitonic basis ($b=x$), the preferred basis ($b=p$), or any other basis in the single-excitation manifold. Equation~\eqref{Eq:S_WNM_matrix_elements} is in the form in which the rotational average can be straightforwardly performed with the final result~\cite{NewJPhys.14.023018}
\begin{equation}
\label{Eq:S_WNM_matrix_elements_rot_averaged}
\begin{split}
    \left\langle b_k\left|S_\mathbf{n}^\mathrm{WNM}\right|b_j\right\rangle^{(\mathrm{avg})}=&\delta_{\mathbf{n},\mathbf{0}}\frac{2I_0\gamma\tau_c}{(\hbar\gamma)^2}\frac{1}{3}\sum_{k'j'}\left(\mathbf{d}_{k'}\cdot\mathbf{d}_{j'}\right)\\&\times\langle b_k|l_{k'}\rangle\langle l_{j'}|b_j\rangle\\
    =&\delta_{\mathbf{n},\mathbf{0}}\frac{2I_0\gamma\tau_c}{(\hbar\gamma)^2}\frac{1}{3}\mathbf{d}_{b_k}\cdot\mathbf{d}_{b_j}^*,
\end{split}
\end{equation}
where $\mathbf{d}_{b_k}=\sum_{k'}\mathbf{d}_{k'}\langle b_k|l_{k'}\rangle$ is the transition dipole moment of state $|b_k\rangle$. In a typical situation, the system of interest consists of many photosynthetic complexes in solution, and the rotational average is performed over random orientation of individual chromophores' dipole moments with respect to the polarization direction. The source term in Eq.~\eqref{Eq:S_WNM_matrix_elements_rot_averaged} depends only on relative orientations of transition dipole moments, which are known, e.g., for the widely investigated $Q_y$-band excitations of the FMO complex.~\cite{BiophysJ.95.847} Representing Eq.~\eqref{Eq:time-evol-n-2nd-rescaling} in basis $\{|b_j\rangle|j\}$ and using the rotationally averaged source term given in Eq.~\eqref{Eq:S_WNM_matrix_elements_rot_averaged}, we obtain a description of incoherent-light driven EET that exploits both the light incoherence (short $\tau_c$) and experimentally available data on relative orientations of transition dipole moments. Such a description features a much more realistic excitation condition than the one we have employed in the study of model dimer (the selective excitation of a local site).

Our NESS approach can also be used to follow the pathways of light-induced excitations, from the point of their generation, through the chromophore network, to the point of their extraction at the load. To elaborate on this, we compute the $|b_k\rangle\langle b_k|$ matrix element of Eq.~\eqref{Eq:time-evol-n-2nd-rescaling} for the RDM ($\mathbf{n}=\mathbf{0}$) and obtain
\begin{equation}
\label{Eq:local-continuity}
    J^{b_k}_\mathrm{gen}-J^{b_k}_\mathrm{RC}-J^{b_k}_\mathrm{rec}+\sum_{k'(\neq k)}J^{b_k b_{k'}}+J^{b_k}_\mathrm{res}=0.
\end{equation}
In Eq.~\eqref{Eq:local-continuity}, $J^{b_k}_\mathrm{gen}$ is the excitation generation flux into the singly excited state $|b_k\rangle$
\begin{equation}
\label{Eq:gen-flux-local}
    J^{b_k}_\mathrm{gen}=\frac{2}{\hbar\gamma}\mathrm{Im}\left\langle b_k\left|\sigma_{eg,\mathbf{0}}^{ss}\mu_{eg}^\dagger\right| b_k\right\rangle,
\end{equation}
$J^{b_k}_\mathrm{RC}$ is the excitation trapping flux from $|b_k\rangle$
\begin{equation}
    J^{b_k}_\mathrm{RC}=-\gamma^{-1}\left\langle b_k\left|\mathcal{L}_\mathrm{RC}\left[\sigma_{ee,\mathbf{0}}^{ss}\right]\right| b_k\right\rangle,
\end{equation}
$J^{b_k}_\mathrm{rec}$ is the excitation recombination flux from $|b_k\rangle$
\begin{equation}
    J^{b_k}_\mathrm{rec}=-\gamma^{-1}\left\langle b_k\left|\mathcal{L}_\mathrm{rec}\left[\sigma_{ee,\mathbf{0}}^{ss}\right]\right| b_k\right\rangle,
\end{equation}
while $J^{b_k b_{k'}}$ (for $k'\neq k$) is the net flux of excitations that are exchanged between states $|b_k\rangle$ and $|b_{k'}\rangle$
\begin{equation}
\label{Eq:J_k_kp}
\begin{split}
    &J^{b_k b_{k'}}=\frac{2}{\hbar\gamma}\mathrm{Im}\left\{\langle b_k|H_M|b_{k'}\rangle\langle b_{k'}|\sigma_{ee,\mathbf{0}}^{ss}|b_k\rangle\right\}\\
    &+2\sum_{j}\sum_{m=0}^{K-1}\sqrt{\frac{\left|c_{m}\right|}{(\hbar\gamma)^2}}\:\mathrm{Im}
\left\{\langle b_{k'}|l_j\rangle\langle l_j|b_k\rangle\langle b_k|\sigma_{ee,\mathbf{0}_{j,m}^+}^{ss}|b_{k'}\rangle\right\}.
\end{split}    
\end{equation}
The last term in Eq.~\eqref{Eq:local-continuity}, $J^{b_k}_\mathrm{res}$, stems from the residual term $2\Delta\delta(t)$ in the decomposition of the bath correlation function into the optimized exponential series, see Eq.~\eqref{Eq:C-j-in-exp-decay}. This term can, therefore, be made arbitrarily small by explicitly treating a sufficient number $K$ of exponentially decaying terms in Eq.~\eqref{Eq:C-j-in-exp-decay}, and will not be considered in the further discussion. For the sake of completeness, let us also note that, if the light incoherence is exploited on the level of Eqs.~\eqref{Eq:G-tau-incoherent}--\eqref{Eq:S_WNM_matrix_elements_rot_averaged}, Eq.~\eqref{Eq:gen-flux-local} for the generation flux should be replaced by
\begin{equation}
    J^{b_k}_\mathrm{gen}=\frac{2I_0\gamma\tau_c}{(\hbar\gamma)^2}\left|\mathbf{d}_{b_k}\right|^2.
\end{equation}

Similarly to Eq.~\eqref{Eq:continuity}, all the fluxes entering Eq.~\eqref{Eq:local-continuity} are dimensionless and the sign in front of them is determined by the "direction" of the flux ($+/-$ if it leads to an increase/a decrease in the population of state $|b_k\rangle$). One can prove that the (global) continuity equation [Eq.~\eqref{Eq:continuity}] is obtained by adding Eq.~\eqref{Eq:local-continuity} for different states $|b_k\rangle$. Therefore, Eq.~\eqref{Eq:local-continuity} can be regarded as the local continuity equation in the basis $\{|b_k\rangle\}$. Here, the term "local" is in no manner connected to the local basis because Eq.~\eqref{Eq:local-continuity} is formulated in an arbitrary basis $\{|b_j\rangle|j\}$. The local continuity equation establishes the balance (on the level of a single-excitation state) between the excitation generation, trapping, and recombination on the one hand and the excitation flow from the considered state toward other states on the other hand.

We believe that the local continuity equation is a potentially interesting feature of our NESS picture because it enables us to track the steady-state excitation pathways. The excitation flux $J^{b_k b_{k'}}$ satisfies $J^{b_k b_{k'}}=-J^{b_{k'} b_k}$, it is positive when the net excitation flow is directed from $b_{k'}$ to $b_k$, while it is negative when the net excitation flow is directed from $b_k$ to $b_{k'}$. One may, therefore, identify the states in which the generation, trapping, and recombination predominantly occur, and then individuate the pathways along which the excitations travel. Such a discussion can be performed in an arbitrary basis of singly excited states, enabling one to follow the excitation pathways in, e.g., the local, excitonic, or preferred basis.

We conclude this section by discussing the generality of the relationship between the stationary and time-dependent pictures that we established for the model dimer in Sec.~\ref{SSec:short-trapping-times}. This relationship relies on the hierarchy of time scales of the dimer's dynamics under incoherent light that is also introduced in Sec.~\ref{SSec:short-trapping-times}. On the other hand, in a multichromophoric aggregate, the rates of excitation transfer between various states can be of the different orders of magnitude, and the excitation may be trapped in certain states, so that the recombination time scale may also become important. As an example, let us take a single unit of the FMO complex which, despite the fact that its contribution to direct light harvesting is minor, has become a paradigmatic system to discuss quantum effects in biological systems.~\cite{Psencik_Mancal_2017} It is known that the intraunit energy transfer predominantly proceeds on subpicosecond time scales.~\cite{JPhysChemLett.7.1653} The excitation transfer from the considered unit of the FMO complex to the RC typically occurs on a $\sim 20\:\mathrm{ps}$ time scale, while the decay time of the lowest FMO level is around $250\:\mathrm{ps}$.~\cite{NatChem.8.705} We may thus speculate that the hierarchy of time scales introduced in Sec.~\ref{SSec:short-trapping-times} is valid in this example and that the reconstruction of the NESS from the dynamics of the driven, but unloaded system is possible. However, in this multichromophoric example, the value of the efficiency genuinely depends on a complex interplay between the excitation generation, relaxation, dephasing, trapping at the RC and recombination and, as such, it cannot be predicted in advance (as is the case for the dimer model). This is even more pronounced in photosynthetically more relevant situations, in which the excitations initially created in antenna complexes reach the RC after many steps of relatively slow interpigment transfer.~\cite{JChemPhys.152.154101} The application of our NESS formalism to such situations is out of the scope of this paper.

\section{Discussion and Conclusion}
\label{Sec:discuss-conclude}
We have provided a detailed and rigorous theoretical description of the light harvesting by a molecular aggregate under conditions that are representative of photosynthetic light harvesting as it occurs in Nature. The picture established in this work takes into account the excitation generation by means of weak incoherent light, their subsequent relaxation and dephasing, as well as excitation trapping by a load (the RC) and recombination. While the generation, relaxation, and dephasing are described in a (numerically) exact manner, which we have reported in the accompanying paper, the excitation trapping and recombination are included on the level of effective Liouvillians.

This piece of research addresses a recurrent question of the possible relevance of quantum coherent effects (understood in a very broad sense) for the natural light harvesting. Our NESS approach provides a physically transparent definition of the light-harvesting efficiency [Eq.~\eqref{Eq:def-eta}] that is basis-invariant, so that we are in position to embark upon the study of possible coherent enhancements of the efficiency. Recent reports have suggested that these coherent enhancements strongly depend on the basis in which the effective trapping and/or recombination Liouvillians are diagonal. Here, we use the fact that the state of an incoherently driven and loaded aggregate is most naturally represented and studied in the so-called preferred basis of the NESS, in which the steady-state RDM is diagonal. This definition of the preferred basis implies that this basis sublimes the joint effect of excitation generation, relaxation, dephasing, trapping, and recombination. The preferred-basis description of the NESS under driving and load can be seen as an analogue of the description of a system of coupled harmonic oscillators in terms of normal modes. While finding the preferred basis is highly nontrivial, as demonstrated throughout this paper, this concept may shed new light on the debate on the role of coherences in the energy transfer under incoherent light.

We have examined the properties of the preferred basis of the NESS of an incoherently driven and loaded dimer by studying the manners in which it is connected to two widely used representations, namely those employing the excitonic and the local bases. The recombination time scale is, in general, significantly longer than any other time scale in the problem, so that, in the limit of long trapping time, the NESS is very similar to the previously studied excited-state equilibrium of an undriven and unloaded system. We also find that the NESS under driving and load carries information that is encoded in the temporal evolution of the unloaded system driven by suddenly turned-on incoherent light. If the radiation is abruptly turned on at $t=0$, the properties of the NESS which arises due to the excitation trapping with time constant $\tau_\mathrm{RC}$ can be quite reliably extracted from the RDM at $t\sim\tau_\mathrm{RC}$. We conclude that the trapping time scales for which such a relation between the NESS and the dynamics of the driven but unloaded system is sensible are basically determined by the time scales of decoherence and relaxation, which are accessible in ultrafast spectroscopy experiments. Since realistic trapping times are in general much longer than decoherence and relaxation time scales, the relation we found between the steady-state and time-dependent pictures is quite general.

We again note that our theoretical and computation approach to obtain the NESS under incoherent driving is general and not limited to the model dimer studied here. We opted for the dimer because the relationships between basis vectors of the preferred basis and the excitonic or local basis can be parameterized by only two real parameters, which have certain physical significance and whose dependence on model parameters can be studied in a systematic manner. In the case of a more complex excitonic aggregate, one should resort to more involved parameterizations of unitary matrices.

\section*{Supplementary Material}
See supplementary material for: (a) the derivation of equations for the transformation parameters $\theta$ and $\Delta$; (b) detailed numerical procedure to compute the nonequilibrium steady state; (c) analysis of the dynamics initiated by a delta-like photoexcitation; (d) the comparison of transformation parameters $\theta_{pl}$ and $\Delta_{pl}$ obtained from stationary and time-dependent pictures in the case of fast trapping.

\begin{acknowledgments}
 The initial stages of this work were supported by Charles University Research Center program No. UNCE/SCI/010 and by Czech Science Foundation (GA\v CR) grant No. 18-18022S. The final stages of this work were supported by the Institute of Physics Belgrade, through the grant by the Ministry of Education, Science, and Technological Development of the Republic of Serbia. Computational resources were supplied by the project "e-Infrastruktura CZ" (e-INFRA LM2018140) provided within the program Projects of Large Research, Development and Innovations Infrastructures. Numerical computations were also performed on the PARADOX supercomputing facility at the Scientific Computing Laboratory of the Institute of Physics Belgrade.
\end{acknowledgments}

\section*{Author's Contributions}
T.M. gave the initial impetus for this work, which started during V.J.'s stay in Prague. V.J. developed the methodology, conducted all numerical computations, analyzed their results, and prepared the initial version of the manuscript. Both authors contributed to the submitted version of the manuscript.

\section*{Data Availability Statement}
The data that support the findings of this study are available from the corresponding authors upon reasonable request.

\newpage
\bibliography{apssamp}

\end{document}


\title{Supplementary Material for:\\
Nonequilibrium steady-state picture of photosynthetic light harvesting}

\author{Veljko Jankovi\'{c}}
\email{veljko.jankovic@ipb.ac.rs}
\affiliation{Scientific Computing Laboratory, Center for the Study of Complex Systems, Institute of Physics Belgrade, University of Belgrade, Pregrevica 118, 11080 Belgrade, Serbia}
\author{Tom\'{a}\v{s} Man\v{c}al}%
 \email{mancal@karlov.mff.cuni.cz}
\affiliation{
 Faculty of Mathematics and Physics, Charles University, Ke Karlovu 5, 121 16 Prague 2, Czech Republic
}

\maketitle
\section{Derivation of Expressions for Transformation Parameters $\theta$ and $\Delta$}
Here, we discuss in greater detail the expressions for the transformation parameters $\theta_{px}$ and $\Delta_{px}$ between the exciton basis and the preferred basis of the NESS. The discussion in the case of localized basis is analogous.

The normalized RDM in the $ee$ sector, $\widetilde{\rho}_{ee}^{ss}$, is expressed in the preferred basis of the NESS and the exciton basis as follows
\begin{equation}
\label{Eq:normalized-rho-general}
 \widetilde{\rho}_{ee}^{ss}=\sum_i\widetilde{p}_i\:|p_i\rangle\langle p_i|=\sum_{jk}\left(\sum_i\langle x_j|p_i\rangle\widetilde{p}_i\langle p_i|x_k\rangle\right)|x_j\rangle\langle x_k|.
\end{equation}
On the other hand, the expression for $\widetilde{\rho}_{ee}^{ss}$ in terms of Pauli matrices
\begin{equation}
\label{Eq:pauli-1}
 \sigma_1=|x_0\rangle\langle x_1|+|x_1\rangle\langle x_0|,\quad\sigma_2=-\im\left(|x_0\rangle\langle x_1|-|x_1\rangle\langle x_0|\right),\quad\sigma_3=|x_0\rangle\langle x_0|-|x_1\rangle\langle x_1|
\end{equation}
reads as
\begin{equation}
\label{Eq:normalized-rho-pauli}
 \widetilde{\rho}_{ee}^{ss}=\frac{1}{2}\left((1+a_3^x)|x_0\rangle\langle x_0|+(a_1^x-\im a_2^x)|x_0\rangle\langle x_1|+(a_1^x+\im a_2^x)|x_1\rangle\langle x_0|+(1-a_3^x)|x_1\rangle\langle x_1|\right).
\end{equation}
Using Eqs.~\eqref{Eq:normalized-rho-general} and~\eqref{Eq:normalized-rho-pauli} together with Eq.~(23) of the main text we obtain
\begin{equation}
\label{Eq:a3-theta}
 a_3^x=(2\widetilde{p}_0-1)\cos(2\theta_{px})=(1-2\widetilde{p}_1)\cos(2\theta_{px})
\end{equation}
\begin{equation}
 a_1^x-\im a_2^x=(2\widetilde{p}_0-1)\:\e^{\im 2\Delta_{px}}\sin(2\theta_{px}).
\end{equation}
The relationships between the Bloch angles $\theta_B^x$ and $\phi_B^x$ and transformation parameters $\theta_{px}$ and $\Delta_{px}$ that presented in the main body of the manuscript now become apparent.

Let us now discuss the range in which $\theta_{px}$ and $\Delta_{px}$ may always be chosen. The Pauli matrices may be chosen as in Eq.~\eqref{Eq:pauli-1} and we this choice will be termed choice 1. There is, however, choice 2, in which $|x_0\rangle$ and $|x_1\rangle$ are permuted
\begin{equation}
 \label{Eq:pauli-2}
 \sigma_1=|x_1\rangle\langle x_0|+|x_0\rangle\langle x_1|,\quad\sigma_2=-\im\left(|x_1\rangle\langle x_0|-|x_0\rangle\langle x_1|\right),\quad\sigma_3=|x_1\rangle\langle x_1|-|x_0\rangle\langle x_0|.
\end{equation}
Then, $a_3^{x,(1)}=-a_3^{x,(2)}$, $a_2^{x,(1)}=-a_2^{x,(2)}$, and $a_1^{x,(1)}=a_1^{x,(2)}$, so that
\begin{equation}
 \cos(2\theta_{px}^{(1)})+\cos(2\theta_{px}^{(2)})=2\cos(\theta_{px}^{(1)}+\theta_{px}^{(2)})\cos(\theta_{px}^{(1)}-\theta_{px}^{(2)})=0,
\end{equation}
\begin{equation}
 \tan(2\Delta_{px}^{(1)})+\tan(2\Delta_{px}^{(2)})=\left(1-\tan(2\Delta_{px}^{(1)})\tan(2\Delta_{px}^{(2)})\right)\tan(2\Delta_{px}^{(1)}+2\Delta_{px}^{(2)})=0.
\end{equation}
It then follows that ($k$ is an integer)
\begin{equation}
\label{Eq:theta-delta-2-1}
 \theta_{px}^{(2)}=\frac{\pi}{2}\pm\theta_{px}^{(1)}+k\pi,\quad \Delta_{px}^{(2)}=-\Delta_{px}^{(1)}+k\frac{\pi}{2}.
\end{equation}
From Eq.~\eqref{Eq:a3-theta}, we know that $\theta_{px}^{(1)}\in(0,\pi/2)$, so that the rotation angle $\theta_{px}^{(2)}$ can always be chosen so than $\theta_{px}^{(2)}\in(0,\pi/4)$. Equation~\eqref{Eq:theta-delta-2-1} suggests that such a choice for $\theta_{px}$ may result in the phase $\Delta_{px}$ acquiring an additional minus sign, which, however, does not affect the range $(-\pi/4,\pi/4)$ of possible values for $\Delta_{px}$. It is for this reason that in Figs.~2--7 of the main text we plot the magnitude $|\Delta_{px}|$.

\section{Computation of the Nonequilibrium Steady State}
As mentioned in the main body of the manuscript, the computational algorithm to obtain the NESS leans on the method proposed in Ref.~\onlinecite{JChemPhys.147.044105} to compute the excited-state equilibrium of a molecular aggregate. The method exploits the fact that the computation of the HEOM steady state can be seen as solving a system of linear algebraic equations, which can be done in an iterative way using, e.g., the Jacobi iteration method. However, the Jacobi iteration method relies on the diagonal dominance of the system, an assumption that is, in general, not satisfied in our problem, especially in the regimes of intermediate and strong system--bath coupling. Equations~(10) and~(11) of the main text are the basis for the following iterative procedure to compute the ADMs $\sigma_{eg,\mathbf{n}}^{ss,\mathrm{new}}$ and $\sigma_{ee,\mathbf{n}}^{ss,\mathrm{new}}$ in the current iteration using the ADMs $\sigma_{eg,\mathbf{n}}^{ss,\mathrm{old}}$ and $\sigma_{ee,\mathbf{n}}^{ss,\mathrm{old}}$ from the previous iteration
\begin{equation}
\label{Eq:time-evol-y-2nd-rescaling}
\begin{split}
 \left(\im\frac{\omega_x-\omega_c}{\gamma}+\frac{\gamma_\mathbf{n}}{\gamma}+(\tau_c\gamma)^{-1}+\epsilon\right)\left\langle x\left|\sigma_{eg,\mathbf{n}}^{ss,\mathrm{new}}\right|g\right\rangle&=\epsilon\left\langle x\left|\sigma_{eg,\mathbf{n}}^{ss,\mathrm{old}}\right|g\right\rangle-\frac{\Delta}{\hbar^2\gamma}\sum_{j}\left\langle x\left|V_j\sigma_{eg,\mathbf{n}}^{ss,\mathrm{old}}\right|g\right\rangle\\
 &+\delta_{\mathbf{n},\mathbf{0}}\frac{\im}{\hbar\gamma}I_{0}\langle x|\mu_{eg}|g\rangle\\
 &+\im\sum_{j}\sum_{m=0}^{K-1}\sqrt{1+n_{j,m}}\sqrt{\frac{\left|c_{m}\right|}{(\hbar\gamma)^2}}
 \left\langle x\left|V_j\sigma_{eg,\mathbf{n}_{j,m}^+}^{ss,\mathrm{old}}\right|g\right\rangle\\
 &+\im\sum_{j}\sum_{m=0}^{K-1}\sqrt{n_{j,m}}\frac{c_{m}/(\hbar\gamma)^2}{\sqrt{\left|c_{m}\right|/(\hbar\gamma)^2}}\left\langle x\left|V_j\sigma_{eg,\mathbf{n}_{j,m}^-}^{ss,\mathrm{old}}\right|g\right\rangle
\end{split}
\end{equation}

\begin{equation}
\label{Eq:time-evol-n-2nd-rescaling}
 \begin{split}
 \left(\im\frac{\omega_x-\omega_{\bar x}}{\gamma}+\frac{\gamma_\mathbf{n}}{\gamma}+\epsilon\right)\left\langle x\left|\sigma_{ee,\mathbf{n}}^{ss,\mathrm{new}}\right|\bar x\right\rangle&=\epsilon\left\langle x\left|\sigma_{ee,\mathbf{n}}^{ss,\mathrm{old}}\right|\bar x\right\rangle-\frac{\Delta}{\hbar^2\gamma}\sum_{j}\left\langle x\left|V_j^\times V_j^\times\sigma_{ee,\mathbf{n}}^{ss,\mathrm{old}}\right|\bar x\right\rangle\\
  &+\frac{\im}{\hbar\gamma}\langle x|\mu_{eg}|g\rangle\left\langle\bar x\left|\sigma_{eg,\mathbf{n}}^{ss,\mathrm{old}}\right|g\right\rangle^*-
  \frac{\im}{\hbar\gamma}\left\langle x\left|\sigma_{eg,\mathbf{n}}^{ss,\mathrm{old}}\right|g\right\rangle\langle\bar x|\mu_{eg}|g\rangle^*\\
  &+\gamma^{-1}\left\langle x\left|\mathcal{L}_\mathrm{rec}[\sigma_{ee,\mathbf{n}}^{ss,\mathrm{old}}]\right|\bar x\right\rangle+\gamma^{-1}\left\langle x\left|\mathcal{L}_\mathrm{RC}[\sigma_{ee,\mathbf{n}}^{ss,\mathrm{old}}]\right|\bar x\right\rangle\\
  &+\im\sum_{j}\sum_{m=0}^{K-1}\sqrt{1+n_{j,m}}\sqrt{\frac{\left|c_{m}\right|}{(\hbar\gamma)^2}}\left\langle x\left|V_j^\times\sigma_{ee,\mathbf{n}_{j,m}^+}^{ss,\mathrm{old}}\right|\bar x\right\rangle\\
  &+\im\sum_{j}\sum_{m=0}^{K-1}\sqrt{n_{j,m}}\frac{c_{m}/(\hbar\gamma)^2}{\sqrt{\left|c_{m}\right|/(\hbar\gamma)^2}}\left\langle x\left|V_j\sigma_{ee,\mathbf{n}_{j,m}^-}^{ss,\mathrm{old}}\right|\bar x\right\rangle\\
  &-\im\sum_j\sum_{m=0}^{K-1}\sqrt{n_{j,m}}
  \frac{c_{m}^*/(\hbar\gamma)^2}{\sqrt{\left|c_{m}\right|/(\hbar\gamma)^2}}\left\langle x\left|\sigma_{ee,\mathbf{n}_{j,m}^-}^{ss,\mathrm{old}}V_j\right|\bar x\right\rangle
 \end{split}
\end{equation}

In Eqs.~\eqref{Eq:time-evol-y-2nd-rescaling} and~\eqref{Eq:time-evol-n-2nd-rescaling}, $|x\rangle$ and $|\bar x\rangle$ are exciton states, $\hbar\omega_x$ and $\hbar\omega_{\bar x}$ are their respective vertical excitation energies, while $\epsilon$ is an adjustable parameter whose value should be tuned so that the steady-state HEOM becomes a diagonally dominant system of linear algebraic equations. The value of $\epsilon$ should be chosen so as to reach a compromise between algorithm stability and numerical accuracy (large $\epsilon$) on the one hand and numerical effort (small $\epsilon$) on the other hand.

Equations~\eqref{Eq:time-evol-y-2nd-rescaling} and~\eqref{Eq:time-evol-n-2nd-rescaling} are solved in the exciton basis because their free-evolution parts are diagonal in that basis, see the c-numbers that multiply the ADM elements in the current iteration on the left-hand sides of these equations. When the trapping and/or recombination Liouvillians are known to be diagonal in the exciton basis, they may be treated in the same manner as the free-evolution terms, which would lead to a more complicated form of the c-numbers appearing on the left-hand sides of Eqs.~\eqref{Eq:time-evol-y-2nd-rescaling} and~\eqref{Eq:time-evol-n-2nd-rescaling}.

The iterative procedure is terminated once the continuity equation [Eq.(...) of the main text] is satisfied with the desired numerical accuracy $\delta$. In more detail, we use the following termination criterion
\begin{equation}
\label{Eq:terminate}
 \frac{\left|J_\mathrm{gen}-J_\mathrm{RC}-J_\mathrm{rec}\right|}{\min\left\{J_\mathrm{gen},J_\mathrm{RC},J_\mathrm{rec}\right\}}<\delta.
\end{equation}
Our numerical computations suggest that the quantity on the left-hand side of Eq.~\eqref{Eq:terminate} monotonously decreases as the algorithm proceeds, so that the termination criterion is sensible. We also monitor changes the ADM elements undergo upon one iteration of the algorithm by following the changes in the following quantities
\begin{equation}
 E_{ee}=\max_{\mathbf{n},\bar x,x}\left\{\left[f(\mathbf{n})\right]^{-1}\left|\left\langle x\left|\sigma_{ee,\mathbf{n}}^{ss,\mathrm{new}}\right|\bar x\right\rangle-\left\langle x\left|\sigma_{ee,\mathbf{n}}^{ss,\mathrm{old}}\right|\bar x\right\rangle\right|\right\},
\end{equation}
\begin{equation}
 E_{eg}=\max_{\mathbf{n},\bar x,x}\left\{\left[f(\mathbf{n})\right]^{-1}\left|\left\langle x\left|\sigma_{eg,\mathbf{n}}^{ss,\mathrm{new}}\right|\bar x\right\rangle-\left\langle x\left|\sigma_{eg,\mathbf{n}}^{ss,\mathrm{old}}\right|\bar x\right\rangle\right|\right\},
\end{equation}
where the rescaling factor $f(\mathbf{n})$ reads as~\cite{JChemPhys.130.084105}
\begin{equation}
 f(\mathbf{n})=\prod_j\prod_{m=0}^{K-1}\left[\left(\frac{|c_m|}{(\hbar\gamma)^2}\right)^{n_{j,m}}n_{j,m}!\right]^{-1/2}.
\end{equation}
We observe that the quantities $E_{ee}$ and $E_{eg}$ monotonically decrease during the course of the algorithm, another sign that our procedure for determining the NESS should lead to correct results.

Another important ingredient of the algorithm is the initial guess for the iterative procedure embodied in Eqs.~\eqref{Eq:time-evol-y-2nd-rescaling} and~\eqref{Eq:time-evol-n-2nd-rescaling}. In Ref.~\onlinecite{JChemPhys.147.044105}, which dealt with the excited-state equilibrium, the initial condition was the purely electronic density matrix in the absence of the environment, i.e., $e^{-\beta H_M}/\mathrm{Tr}_M\left\{e^{-\beta H_M}\right\}$. Here, however, we have incoherent driving, trapping, and recombination, so that a natural initial guess for the NESS can be obtained by solving the corresponding Redfield equation. In our companion paper, we presented the derivation of the Redfield equation under driving.~\cite{technical-vj-tm} The appropriate modifications to take into account excitation trapping and recombination are described in the main body of the manuscript. The corresponding steady-state Redfield equations in the $eg$ and $ee$ sectors read as
\begin{equation}
\label{Eq:ss-red-y}
\begin{split}
0=-\im\left(\frac{\omega_x-\omega_c}{\gamma}-\im(\tau_c\gamma)^{-1}\right)\langle x|\rho_{eg}^{ss}|g\rangle+\frac{\im}{\hbar\gamma}I_0\langle x|\mu_{eg}|g\rangle-\gamma^{-1}\sum_{x'}\left(\sum_{\tilde x}\mathrm{Re}\:\Gamma_{x\tilde x\tilde x x'}\right)\langle x'|\rho_{eg}^{ss}|g\rangle,
\end{split}
\end{equation}
\begin{equation}
\label{Eq:ss-red-n}
\begin{split}
0&=-\im\frac{\omega_x-\omega_{\bar x}}{\gamma}\langle x|\rho_{ee}^{ss}|\bar x\rangle+\frac{\im}{\hbar\gamma}\langle\bar x|\rho_{eg}^{ss}|g\rangle^*\langle x|\mu_{eg}|g\rangle-\frac{\im}{\hbar\gamma}\langle\bar x|\mu_{eg}|g\rangle^*\langle x|\rho_{eg}^{ss}|g\rangle\\
&+\gamma^{-1}\sum_{\bar x'x'}\left(\mathrm{Re}\:\Gamma_{\bar x'\bar x x x'}+\mathrm{Re}\:\Gamma_{x'x\bar x\bar x'}^*
-\delta_{\bar x'\bar x}\sum_{\tilde x}\mathrm{Re}\:\Gamma_{x\tilde x\tilde xx'}-\delta_{x'x}\sum_{\tilde x}\mathrm{Re}\:\Gamma_{\bar x\tilde x\tilde x\bar x'}^*\right)\langle x'|\rho_{ee}^{ss}|\bar x'\rangle\\
&+\gamma^{-1}\left\langle x\left|\mathcal{L}_\mathrm{rec}[\rho_{ee}^{ss}]\right|\bar x\right\rangle+\gamma^{-1}\left\langle x\left|\mathcal{L}_\mathrm{RC}[\rho_{ee}^{ss}]\right|\bar x\right\rangle,
\end{split}
\end{equation}
where the tetradic quantity $\Gamma_{\bar xx\bar x' x'}$ is
\begin{equation}
\label{Eq:gamma}
\Gamma_{\bar x x\bar x' x'}=\sum_{j}\langle\bar x|j\rangle\langle j|x\rangle\langle\bar x'|j\rangle\langle j|x'\rangle
\int_0^{+\infty}\dif s\:\frac{C(s)}{\hbar^2}\exp\left(\im(\omega_{x'}-\omega_{\bar x'})s\right).
\end{equation}
In Eqs.~\eqref{Eq:ss-red-y} and~\eqref{Eq:ss-red-n}, we follow a common practice and neglect imaginary parts of $\Gamma_{\bar xx\bar x' x'}$. For the Drude--Lorentz spectral density, and under the assumption of purely real exciton wave functions $\langle j|x\rangle$, the corresponding real parts can be computed analytically to yield
\begin{equation}
\label{Eq:Re-gamma}
 \mathrm{Re}\:\Gamma_{\bar x x\bar x' x'}=\left[\sum_{j}\langle\bar x|j\rangle\langle j|x\rangle\langle\bar x'|j\rangle\langle j|x'\rangle\right]\mathcal{C}(\omega_{x'}-\omega_{\bar x'}),
\end{equation}
where
\begin{equation}
\label{Eq:cal-C}
\mathcal{C}(\omega)=\mathrm{Re}\int_0^{+\infty}\dif s\:C(s)\:\e^{\im\omega s}=\begin{cases}
\frac{1}{2}\times\frac{2\pi}{\hbar}\times\left(\frac{2}{\pi}\lambda\frac{\omega\gamma}{\omega^2+\gamma^2}\right)\left(1+n_\mathrm{BE}(\omega)\right),&\omega>0;\\
\frac{1}{2}\times\frac{2\pi}{\hbar}\times\left(\frac{2}{\pi}\lambda\frac{|\omega|\gamma}{|\omega|^2+\gamma^2}\right)n_\mathrm{BE}(|\omega|),&\omega<0;\\
2\times\frac{\lambda}{\hbar\gamma}\times\frac{k_BT}{\hbar},&\omega=0.
\end{cases}
\end{equation}
Upon solving Eqs.~\eqref{Eq:ss-red-y} and~\eqref{Eq:ss-red-n}, we explicitly check that the continuity equation $J_\mathrm{gen}-J_\mathrm{rec}-J_\mathrm{RC}=0$
is satisfied.

The convergence of the HEOM method should always be checked against the maximal depth of the hierarchy and the number of terms in the optimized exponential series for the bath correlation function. Let us first concentrate on the convergence with respect to the depth of the hierarchy. We gradually increase the depth of the hierarchy in the following manner:
\begin{enumerate}
 \item we start with Eqs.~\eqref{Eq:time-evol-y-2nd-rescaling} and~\eqref{Eq:time-evol-n-2nd-rescaling} up to depth $D=2$; the initial guess for the RDM is obtained by solving Eqs.~\eqref{Eq:ss-red-y} and~\eqref{Eq:ss-red-n}, while the ADMs on depths 1 and 2 are set to zero; the numerical accuracy with which the continuity equation is satisfied is set to $\delta_2$, see Eq.~\eqref{Eq:terminate};
 \item we use the solution up to depth $D\geq 2$ as the initial guess for the computations up to depth $D+2$ (the ADMs at depths $D+1$ and $D+2$ are set to zero); the numerical accuracy with which the continuity equation is satisfied in the computation up to depth $D+2$ is $\delta_{D+2}=c\cdot\delta_D$, where $c<1$ (if the numerical accuracy is not downscaled, the algorithm at depth $D+2$ terminates immediately). 
\end{enumerate}
In this manner, we are able to check how the quantities of our interest, in particular transformation parameters between exciton/localized basis and the preferred basis of the NESS, depend on the maximal depth of the hierarchy. For the values of model parameters summarized in Table I of the main text, the deepest hierarchy is constructed for the largest reorganization energy ($\lambda=400\:\mathrm{cm}^{-1}$), and its depth is 14.

We now briefly discuss the convergence of the HEOM with respect to $K$, and concentrate on the values of model parameters listed in Table I of the main text. These values satisfy the low-temperature approximation $\beta\hbar\gamma\ll 1$ reasonably well. When the interaction with the environment is weak, the steady-state Redfield equations [Eqs.~\eqref{Eq:ss-red-y} and~\eqref{Eq:ss-red-n}] should present a good description of the situation of our interest. From Eqs.~\eqref{Eq:gamma}--\eqref{Eq:cal-C}, we see that, in this case, the relaxation tensor depends on the full spectral density, so that we should have $K>1$. We have checked that $K=3$ is a reasonable choice for the weak coupling to the environment. On the other hand, for stronger excitation--environment coupling, we have numerically verified that it is enough to take $K=1$. 

\newpage

\section{Analysis of the Dynamics Initiated by a $\delta$-like Photoexcitation}
\begin{figure}[htbp]
 \includegraphics{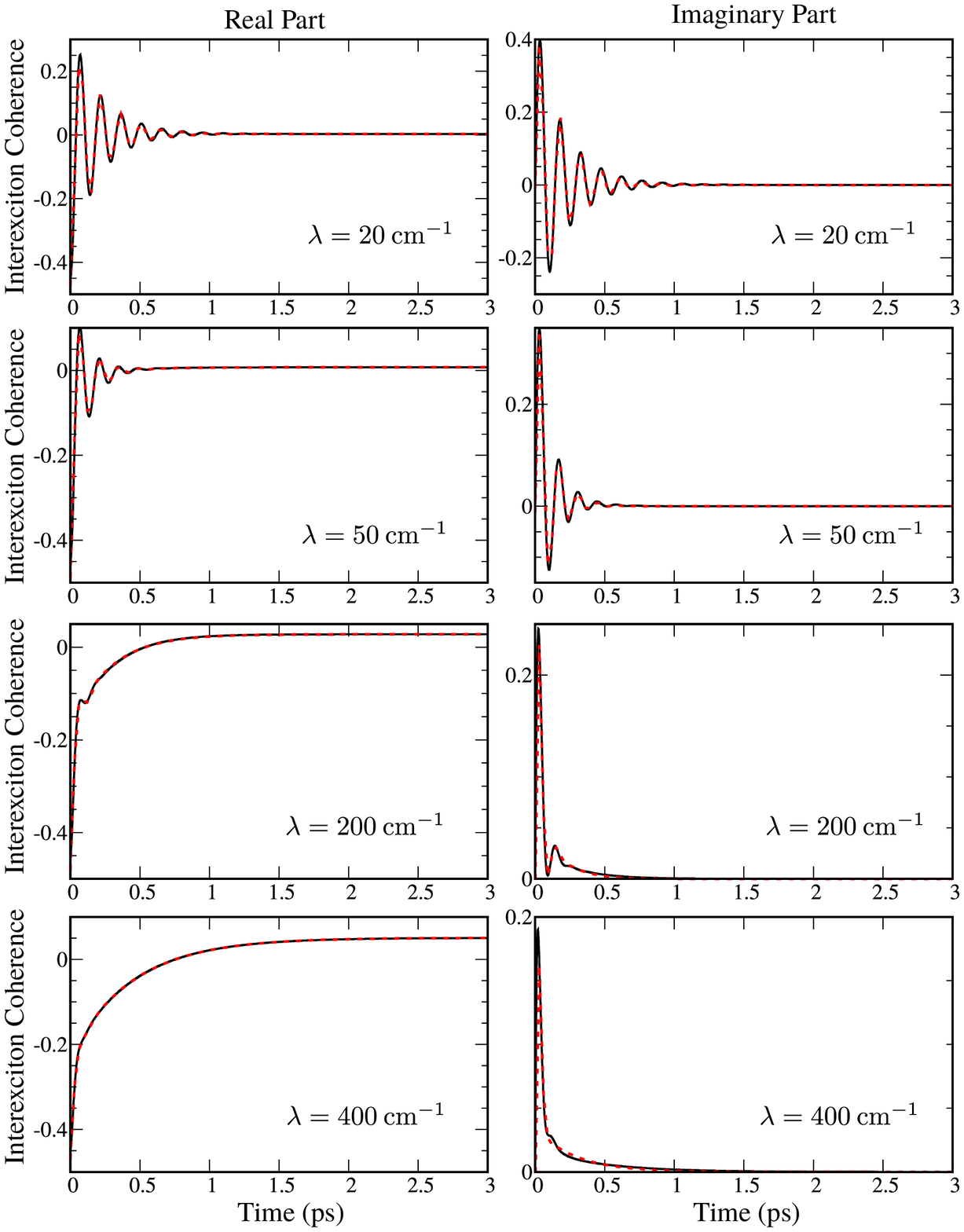}
 \caption{Time dependence of the real (left column) and imaginary (right column) part of the interexciton coherence following a sudden $\delta$-like excitation of the model dimer. Solid lines are obtained by propagating HEOM, while dashed lines are best fits to numerical data using the fitting functions given in Eqs.~\eqref{Eq:fit-re} and~\eqref{Eq:fit-im}. The best fitting parameters are summarized in Tables~\ref{Tab:fitting_params_20}--\ref{Tab:fitting_params_400}.}
 \label{Fig:delta-time-scales}
\end{figure}

Here, we analyze in greater detail the dynamics of the model dimer initiated by an impulsive photoexcitation and extract the time scales of such dynamics. In particular, we examine time dependence of the real and imaginary part of the interexciton coherence on a picosecond time scale following a sudden $\delta$-like excitation at $t=0$. The real part of the interexciton coherence is fitted using
\begin{equation}
\label{Eq:fit-re}
 \mathrm{Re}\left\{\rho_{01}(t)\right\}=a_0+a_1\:e^{-t/a_2}\cos(a_3 t)+a_4\:e^{-t/a_5}+a_6\:e^{-t/a_7},
\end{equation}
while the fitting function for the imaginary part reads as
\begin{equation}
\label{Eq:fit-im}
 \mathrm{Im}\left\{\rho_{01}(t)\right\}=b_1\:e^{-t/b_2}\sin(b_3 t)+b_4\:e^{-t/b_5}+b_6\:e^{-t/b_7}.
\end{equation}
In Fig.~\ref{Fig:delta-time-scales}, we present the results of computations and fit, while the best values of fitting parameters are summarized in Tables~\ref{Tab:fitting_params_20}--\ref{Tab:fitting_params_400}.

\newpage

\begin{table}[htbp]
 \caption{Fitting Parameters for $\lambda=20\:\mathrm{cm}^{-1}$.}
 \label{Tab:fitting_params_20}
 \centering
 \begin{tabular}{c | c}
  \hline
  Parameter (Unit) & Value\\
  \hline
  $a_0$ (-) & $2.96\times 10^{-3}$\\
  $a_1$ (-) & $-0.3933$\\
  $a_2$ (fs) & 193.9\\
  $\hbar a_3$ (cm$^{-1}$) & 228.5\\
  $a_4$ (-) & $-0.091$\\
  $a_5$ (fs) & 84\\
  $a_6$ (-) & 0.010\\
  $a_7$ (fs) & 300\\
  \hline\hline
  $b_1$ (-) & 0.4460\\
  $b_2$ (fs) & 192.8\\
  $\hbar b_3$ (cm$^{-1}$) & 229.1\\
  $b_4$ (-) & $-0.0277$\\
  $b_5$ (fs) & 103\\
  $b_6$ (-) & 0.0209\\
  $b_7$ (fs) & 282\\
 \end{tabular}
\end{table}

\begin{table}[htbp]
 \caption{Fitting Parameters for $\lambda=50\:\mathrm{cm}^{-1}$.}
 \label{Tab:fitting_params_50}
 \centering
 \begin{tabular}{c | c}
  \hline
  Parameter (Unit) & Value\\
  \hline
  $a_0$ (-) & $7.24\times 10^{-3}$\\
  $a_1$ (-) & $-0.3146$\\
  $a_2$ (fs) & 105.2\\
  $\hbar a_3$ (cm$^{-1}$) & 235.7\\
  $a_4$ (-) & $-0.03135$\\
  $a_5$ (fs) & 277\\
  $a_6$ (-) & $-0.14762$\\
  $a_7$ (fs) & 60.70\\
  \hline\hline
  $b_1$ (-) & 0.43085\\
  $b_2$ (fs) & 99.20\\
  $\hbar b_3$ (cm$^{-1}$) & 240.6\\
  $b_4$ (-) & $-0.06465$\\
  $b_5$ (fs) & 165.2\\
  $b_6$ (-) & 0.0526\\
  $b_7$ (fs) & 148.0\\
 \end{tabular}
\end{table}

\newpage

\begin{table}[htbp]
 \caption{Fitting Parameters for $\lambda=200\:\mathrm{cm}^{-1}$.}
 \label{Tab:fitting_params_200}
 \centering
 \begin{tabular}{c | c}
  \hline
  Parameter (Unit) & Value\\
  \hline
  $a_0$ (-) & 0.02787\\
  $a_1$ (-) & $-0.163$\\
  $a_2$ (fs) & 51.4\\
  $\hbar a_3$ (cm$^{-1}$) & 230\\
  $a_4$ (-) & $-0.141$\\
  $a_5$ (fs) & 43.9\\
  $a_6$ (-) & $-0.2038$\\
  $a_7$ (fs) & 270.7\\
  \hline\hline
  $b_1$ (-) & 0.427\\
  $b_2$ (fs) & 36.9\\
  $\hbar b_3$ (cm$^{-1}$) & 250\\
  $b_4$ (-) & $-0.0754$\\
  $b_5$ (fs) & 15.92\\
  $b_6$ (-) & 0.0662\\
  $b_7$ (fs) & 156.7\\
 \end{tabular}
\end{table}

\begin{table}[htbp]
 \caption{Fitting Parameters for $\lambda=400\:\mathrm{cm}^{-1}$.}
 \label{Tab:fitting_params_400}
 \centering
 \begin{tabular}{c | c}
  \hline
  Parameter (Unit) & Value\\
  \hline
  $a_0$ (-) & 0.050754\\
  $a_1$ (-) & $-0.082$\\
  $a_2$ (fs) & 42.80\\
  $\hbar a_3$ (cm$^{-1}$) & 222\\
  $a_4$ (-) & $-0.163$\\
  $a_5$ (fs) & 44.7\\
  $a_6$ (-) & $-0.2746$\\
  $a_7$ (fs) & 446.5\\
  \hline\hline
  $b_1$ (-) & 0.8\\
  $b_2$ (fs) & 19.5\\
  $\hbar b_3$ (cm$^{-1}$) & 170\\
  $b_4$ (-) & $-0.0496$\\
  $b_5$ (fs) & 16\\
  $b_6$ (-) & 0.0349\\
  $b_7$ (fs) & 277\\
 \end{tabular}
\end{table}

\newpage

\section{Transformation Parameters $\theta_{pl}$ and $\Delta_{pl}$ under Fast Trapping}

\begin{figure}[htbp!]
 \includegraphics[scale=1.05]{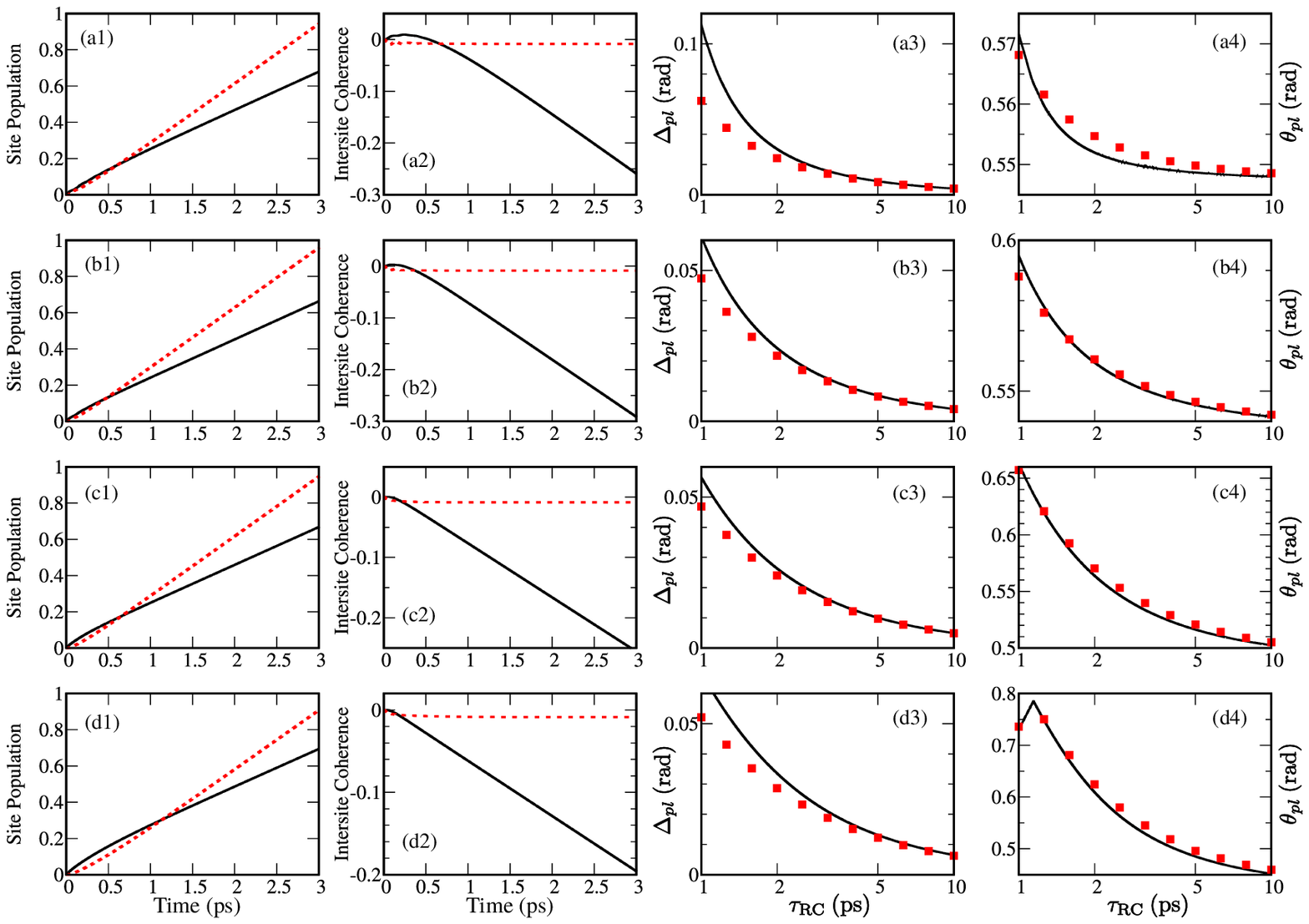}
 \caption{(a1)--(d1): Time dependence of populations of localized states $|l_0\rangle$ (solid line) and $|l_1\rangle$ (dashed line) of the incoherently driven and unloaded model dimer for different values of the reorganization energy. (a2)--(d2): Time dependence of the real (solid line) and imaginary (dashed line) parts of the intersite coherence of the incoherently driven and unloaded model dimer for different values of the reorganization energy. Both site populations and intersite coherences are measured in units of $I_0\:d_{eg}^2/(\hbar\gamma)^2$. The excitation is suddenly turned on at $t=0$. Dependence of the transformation parameters $\Delta_{px}$ [(a3)--(d3)] and $\theta_{px}$ [(a4)--(d4)] between the localized basis and the preferred basis of the NESS on the trapping time constant $\tau_\mathrm{RC}\in(1,10)\:\mathrm{ps}$ for different values of the reorganization energy. Solid lines are computed using time traces of a driven and unloaded model dimer at $t=\tau_\mathrm{RC}$, while squares emerge from the computation of the NESS using Eqs.~(10) and~(11) of the main text. The scale on the abscissa ($\tau_\mathrm{RC}$) in (a3)--(d4) is logarithmic. Trapping at the RC is governed by the localized-trapping Liouvillian [Eq.~(18) of the main text]. The values of the reorganization energy are 20~cm$^{-1}$ [(a1)--(a4)], 50\:cm$^{-1}$ [(b1)--(b4)], 200\:cm$^{-1}$ [(c1)--(c4)], and 400\:cm$^{-1}$ [(d1)--(d4)].}
 \label{Fig:fast}
\end{figure}
\bibliography{apssamp}